\documentclass[11pt]{article}
\usepackage{verbatim}
\usepackage{amsmath,amssymb}
\makeatletter \@addtoreset{equation}{section} \makeatother

\newcommand{\noi}{\vspace{12pt}\noindent}
\newcommand{\beq}{\begin{equation}}
\newcommand{\eeq}{\end{equation}}
\newcommand{\bea}{\begin{eqnarray}}
\newcommand{\eea}{\end{eqnarray}}

\newcommand{\e}[1]{{(\ref{#1})}}
\newcommand{\eq}[1]{{eq.\ (\ref{#1})}}
\newcommand{\es}[2]{{(\ref{#1}) and (\ref{#2})}}
\newcommand{\eqs}[2]{{eqs.\ (\ref{#1}) and (\ref{#2})}}
\newcommand{\Ref}[1]{{Ref.~\cite{#1}}}
\newcommand{\mb}[1]{{\mbox{${#1}$}}}
\newcommand{\equi}[1]{\stackrel{{#1}}{=}}

\newcommand{\C}{\mathbb{C}}
\newcommand{\N}{\mathbb{N}}
\newcommand{\R}{\mathbb{R}}
\newcommand{\Z}{\mathbb{Z}}

\newcommand{\cA}{{\cal A}}
\newcommand{\cC}{{\cal C}}
\newcommand{\cF}{{\cal F}}
\newcommand{\cG}{{\cal G}}
\newcommand{\cL}{{\cal L}}
\newcommand{\cM}{{\cal M}}
\newcommand{\cN}{{\cal N}}
\newcommand{\cU}{{\cal U}}
\newcommand{\ii}{{\tilde{\imath}}}
\newcommand{\jj}{{\tilde{\jmath}}}
\newcommand{\kk}{{\tilde{k}}}
\newcommand{\mm}{{\tilde{m}}}

\newcommand{\ie}{{${ i.e., \ }$}}
\newcommand{\eg}{{${ e.g., \ }$}}

\newcommand{\cf}{{cf.\ }}

\newcommand{\aka}{{also known as }}
\newcommand{\wrt}{{with respect to }}
\newcommand{\wrtt}{{with respect to the }}
\newcommand{\wtho}{{with the help of }}

\newcommand{\lhs}{{left-hand side }}
\newcommand{\rhs}{{right-hand side }}
\renewcommand{\~}{ \ }
\renewcommand{\=}{ \ = \ }

\renewcommand{\Tilde}{\widetilde}

\newcommand{\eps}{\varepsilon^{}}
\newcommand{\p}{\!{}^{}}
\newcommand{\q}{{}^{}}
\newcommand{\dd}{\mathrm{d}}

\newcommand{\Id}{{\rm Id}}
\newcommand{\Mat}{{\rm Mat}}
\newcommand{\End}{{\rm End}}
\newcommand{\Ad}{{\rm Ad}}
\newcommand{\ad}{{\rm ad}}
\newcommand{\tr}{{\rm tr}}

\newcommand{\gauge}{{\rm gauge}}

\newcommand{\Hf}{\frac{1}{2}}

\newcommand{\twobyone}[2]{\left[\begin{array}{c}{#1} \cr
                                {#2} \end{array} \right]}

\newcommand{\twobytwo}[4]{\left[\begin{array}{cc}{#1}&{#2} \cr
                                  {#3} & {#4} \end{array} \right]}
\newcommand{\threebythree}[9]{\left[\begin{array}{ccc}{#1}&{#2}&{#3} \cr
                 {#4}&{#5}&{#6} \cr {#7}&{#8}&{#9} \end{array} \right]}

\newcommand{\twostack}[2]{\begin{array}{c} \lower.8ex\hbox{${#1}$}
                     \cr \raise.8ex\hbox{${#2}$} \end{array}}
\newcommand{\deder}[1]{\frac{ 
 \stackrel{\raise.2ex\hbox{$\leftarrow$}}{\delta^{r}}   } 
 {   \delta {#1}}  }
\newcommand{\dedel}[1]{\frac{ 
 \stackrel{\lower.3ex \hbox{$\rightarrow$}}{\delta^{\ell}}   }
 {   \delta {#1}}  }
\newcommand{\papar}[1]{\frac{  
 \stackrel{\raise.2ex\hbox{$\leftarrow$}}{\partial^{r}}   } 
 {   \partial {#1}}  }
\newcommand{\papal}[1]{\frac{ 
 \stackrel{\lower.3ex \hbox{$\rightarrow$}}{\partial^{\ell}}   }
 {   \partial {#1}}  }
\newcommand{\rpa}[1]{{ 
 \stackrel{\raise.2ex\hbox{$\leftarrow$}}{\partial^{r}_{#1}}   }}
\newcommand{\lpa}[1]{{ 
 \stackrel{\lower.3ex\hbox{$\rightarrow$}}{\partial^{\ell}_{#1}}  }}
\newcommand{\larrow}[1]{\stackrel{\rightarrow}{#1}}

\newcommand{\proofbox}{\begin{flushright}{\hfill \ensuremath{\Box}}
\end{flushright}}

\newtheorem{theorem}{Theorem}[section]
\newtheorem{assumption}[theorem]{Assumption}

\newtheorem{definition}[theorem]{Definition}

\newtheorem{lemma}[theorem]{Lemma}
\newtheorem{observation}[theorem]{Observation}

\newtheorem{proposition}[theorem]{Proposition}

\begin{document}
\thispagestyle{empty}
\title{\Large{\bf A Triplectic Bi-Darboux Theorem\\ 
and Para-Hypercomplex Geometry}}
\author{{\sc Igor~A.~Batalin}$^{a}$ and {\sc Klaus~Bering}$^{b}$ \\~\\
$^{a}$I.E.~Tamm Theory Division\\
P.N.~Lebedev Physics Institute\\Russian Academy of Sciences\\
53 Leninsky Prospect\\Moscow 119991\\Russia\\~\\
$^{b}$Institute for Theoretical Physics \& Astrophysics\\
Masaryk University\\Kotl\'a\v{r}sk\'a 2\\CZ--611 37 Brno\\Czech Republic}
\maketitle
\vfill
\begin{abstract}
We provide necessary and sufficient conditions for a bi-Darboux Theorem on
triplectic manifolds. Here triplectic manifolds are manifolds equipped with 
two compatible, jointly non-degenerate Poisson brackets with mutually
involutive Casimirs, and with ranks equal to $2/3$ of the manifold dimension.
By definition bi-Darboux coordinates are common Darboux coordinates for two
Poisson brackets. We discuss both the Grassmann-even and the Grassmann-odd
Poisson bracket case. Odd triplectic manifolds are, \eg relevant for
$Sp(2)$-symmetric field-antifield formulation. We demonstrate a one-to-one
correspondence between triplectic manifolds and para-hypercomplex manifolds.
Existence of bi-Darboux coordinates on the triplectic side of the 
correspondence translates into a flat Obata connection on the
para-hypercomplex side.
\end{abstract}
\vfill
\begin{quote}
MSC number(s): 37J99; 53D99; 55R10; 57R30; 58C50; 70S05.  
\hrule width 5.cm \vskip 2.mm \noindent 
$^{a}${\small E--mail:~{\tt batalin@lpi.ru}} \hspace{10mm}
$^{b}${\small E--mail:~{\tt bering@physics.muni.cz}} \\
\end{quote}

\newpage
\tableofcontents

\section{Introduction}
\label{secintro}

\noi
A {\em bi-Poisson} supermanifold is a supermanifold equipped with two Poisson
brackets. We shall here discuss both the case of two Grassmann-even Poisson
brackets and the case of two Grassmann-odd Poisson brackets (\aka
antibrackets).

\noi
Compatible Grassmann-even bi-Poisson structures have been studied extensively
for more than thirty years in integrable systems \cite{magri78,gelfand00},
usually with the extra assumption that at least one of the two Poisson
structures are non-degenerate(=symplectic). 

\noi
Compatible Grassmann-odd Poisson structures appear in the $Sp(2)$-symmetric
version \cite{bltla2,bm95,bms95,djb95,bm96} of the field-antifield formulation
\cite{bv81,bv83,bv84}. This quantization scheme naturally live on a
$3n$-dimensional odd triplectic manifold $\cM$. In particular, the total
dimension of the underlying manifold $\cM$ is a multiplum of $3$. 
(In order to be as general as possible, we will here only be interested in the
two antibrackets, and ignore the fact that the $Sp(2)$-symmetric
field-antifield formulation also contains two Grassmann-odd vector fields 
$V^{a}$, $a\in\{1,2\}$, which in turn would force the dimension of $\cM$ to be
a multiplum of $6$ rather than $3$.) {\em Triplectic structures} will in this
paper refer to bi-Poisson structures that are jointly non-degenerate, with
mutually involutive Casimirs, and with $2/3$ ranks, \cf
Definition~\ref{deftriplectic}.

\noi
The main purpose of our paper is to investigate the possible existence of 
bi-Darboux coordinates for triplectic structures, \ie if it is possible
to locally bundle the coordinates of a triplectic manifold $\cM$ into triplets
$(q^{i},p\q_{1i},p\q_{2i})$ of one position variable $q^{i}$ and two momentum
variables $p\q_{1i}$, $p\q_{2i}$ each.
The papers \cite{grisemi97,grisemi98} by Grigoriev and Semikhatov state the 
necessary and sufficient factorization condition \e{factorization} for the
corresponding version of bi-Darboux Theorem, \cf 
Theorem~\ref{theoremfactorization}, although without a complete\footnote{
In detail, the existence of a function $H$ in eq.\ (3.17) of \Ref{grisemi97}
relies implicitly on an un-proven version of the bi-Poincar\'e Lemma, which
is covered in the case $E^{i}\q_{\jj}=\delta^{i}_{\jj}$ by our new
bi-Poincar\'e Lemma~\ref{lemmabipoincarelemma}.} proof. 
We will here give a proof of the bi-Darboux 
Theorem~\ref{theoremfactorization} \wtho a new bi-Poincar\'e
Lemma~\ref{lemmabipoincarelemma}. It turns out that the usual super-proof
technique \cite{bv85} for the standard Poincar\'e Lemma (which at its core is
based on defining a suitable pairing between variables of opposite
Grassmann-parity) is not applicable to the triplectic setting.
Instead we give a proof of the bi-Poincar\'e Lemma~\ref{lemmabipoincarelemma}
\wtho $sl(2,\mathbb{C})$ representation theory.

\noi
The paper is organized as follows. Section~\ref{secpoissonstructure}
basic definitions and establishes notation. The main bi-Darboux
Theorem~\ref{theoremfactorization} is stated in
Subsection~\ref{secbidarbouxthm}, and proved in
Section~\ref{secproofbidarboux02}.
Sections~\ref{secpoincarelemma}--\ref{secproofbidarboux} develop material
and formalism needed in the proof. Section~\ref{secbict} contains a discussion
of bi-canonical transformations, and Section~\ref{secparahypercomplex}
discusses a one-to-one correspondence between triplectic manifolds and
para-hypercomplex manifolds. Para-hypercomplex geometry is a rapidly
developing topic in differential geometry
\cite{swann05,andrada05,alekseevsky08,alekseevsky09} and in twisted
supersymmetric $\cN=(4,4)$ non-linear sigma-models \cite{gotelind11}.
Subsection~\ref{secobata} shows how para-hypercomplex supermanifolds are
endowed with a unique Obata connection \cite{obata56}. It turns out that the 
necessary and sufficient factorization condition \e{factorization} from the
main bi-Darboux Theorem~\ref{theoremfactorization} is equivalent to that the 
Obata connection is flat.
{}Finally, Appendix~\ref{appbipoincarelemma} contains a proof of bi-Poincar\'e
Lemma~\ref{lemmabipoincarelemma}, while Appendix~\ref{appliegroup} lists some
Lie group facts used in Section~\ref{secparahypercomplex}.

\subsection{General Remarks About Notation}
\label{secnotation}

\noi
Adjectives from supermathematics such as ``graded'', ``super'', etc., are
implicitly implied. The sign conventions are such that two exterior forms
$\xi$ and $\eta$, of Grassmann-parity $\eps_{\xi}$, $\eps_{\eta}$
and of form-degree $p\q_{\xi}$, $p\q_{\eta}$, commute in the following
graded sense
\beq 
 \eta \wedge \xi \=
(-1)^{\eps_{\xi}\eps_{\eta}+p\q_{\xi}p\q_{\eta}}\xi\wedge\eta
\label{etawedgexi}
\eeq
inside the exterior algebra. The exterior wedge symbol
``\mb{\wedge}'' is often not written explicitly, as it is redundant
information that can be deduced from the Grassmann- and form-parity. The
commutator \mb{[F,G]} and anticommutator \mb{\{F,G\}^{}_{+}} of two operators
\mb{F} and \mb{G} are 
\bea
[F,G]\~&:=&FG-(-1)^{\eps_{F}\eps_{G}+p\q_{F}p\q_{G}}GF\~,\label{com01} \\
\{F,G\}^{}_{+}&:=&FG+(-1)^{\eps_{F}\eps_{G}+p\q_{F}p\q_{G}}GF\~.\label{anticom01}
\eea
Note that in Section~\ref{secpoincarelemma},
Subsection~\ref{secbipoincarelemma}, and Appendix~\ref{appbipoincarelemma},
there appear some objects $\eta\q_{i}$, $x^{i}_{3}$, etc., which are
semantically referred to as ``forms'', although we will actually not assign any
non-zero form-degree $p$ to them that affects their commutation properties 
\e{etawedgexi}.

\section{Bi-Poisson Structure}
\label{secpoissonstructure}

\subsection{Poisson Pencil}
\label{secpoissonpencil}

\noi
Let there be given a manifold $\cM$ of dimension
$3n$ with two compatible Poisson brackets $\{\cdot,\cdot\}^{a}$, $a\in\{1,2\}$,
of rank $2n$, with common intrinsic Grassmann parity $\eps$, 
\beq
\eps(\{f,g\}^{a})\=\eps_{f}+\eps+\eps_{g}\~, \qquad f,g \in C^{\infty}(\cM)\~,
\qquad a\in\{1,2\}\~,
\label{grassmann01}
\eeq
and with symmetry
\beq
\{f,g\}^{a}\= -(-1)^{(\eps_{f}+\eps)(\eps_{g}+\eps)}\{g,f\}^{a}\~, 
\qquad f,g \in C^{\infty}(\cM)\~, \qquad a\in\{1,2\}\~.\label{sym01}
\eeq
In other words, the case $\eps\!=\!0$ ($\eps\!=\!1$) corresponds to a pair 
of even (odd) Poisson brackets, respectively.
The word {\em compatible} means that any $\R$-linear combination of the
two Poisson brackets $\{\cdot,\cdot\}^{a}$, $a\in\{1,2\}$, is again a Poisson
bracket, \cf Subsection~\ref{secglobalgl2}.
Alternatively, one says that the two Poisson structures form a 
{\em Poisson pencil}. In particular, the two Poisson brackets satisfy a 
{\em symmetrized} Jacobi identity  
\beq
\sum_{{\rm cycl.}~f,g,h}(-1)^{(\eps_{f}+\eps)(\eps_{h}+\eps)}
\{\{f,g\}^{\{a},h\}^{b\}} \= 0\~, \qquad f,g,h \in C^{\infty}(\cM)\~,
\qquad a,b\in\{1,2\}\~, \label{mixjacid}
\eeq
which contains the Jacobi identity for each Poisson brackets, and a six-term
{\em mixed} Jacobi identity.

\noi
The symmetrized Jacobi identity \e{mixjacid} is a very important geometrical 
input. A good part of the following 
Sections~\ref{secpoissonstructure}--\ref{secpoincarelemma} will deal with
extracting exhaustively the huge amount of geometric information that it 
contains.

\subsection{Global $GL(2)$ Covariance}
\label{secglobalgl2}

\noi
The construction must behave covariantly under the group\footnote{The matrix 
$g\q_{a}{}^{b}$ for the group element $g\in GL(2)$ is unconventionally 
written with its indices upside-down. {}For instance, the transposed matrix is
written as $(g^T)^{a}\q_{b}:= g\q_{b}{}^{a}$.} 
$GL(2)=SL(2)\times\R^{\times}$ of global rotations of the two Poisson brackets,
\beq
\{\cdot,\cdot\}^{a}\~\to\~\{\cdot,\cdot\}^{\prime b}
\=\{\cdot,\cdot\}^{a}\~ (g^{-1})\q_{a}{}^{b}\~, \qquad g\in GL(2)\~. 
\label{gl2rotationpb}
\eeq
where the group $GL(2)$ by definition acts from left.
It turns out that the overall scaling group\footnote{
The scaling group $\R^{\times}$ is absent in the $Sp(2)$-symmetric 
field-antifield formulation \cite{bltla2,bm95,bms95,djb95,bm96} 
because of explicit appearances of the Levi-Civita $\epsilon^{ab}$ tensor.
See also Appendix~\ref{appliegroup}.} 
$\R^{\times}\equiv \R\backslash \{0\}$ acts trivially
(basically because it belongs to the center of $GL(2)$),
so that only the $SL(2)=Sp(2)$ part is interesting. We should stress that we
here do {\em not} a priori assume the existence of an ``intrinsic'' group
action ``.'' :$SL(2)\times\cM \to \cM$ on the manifold $\cM$, and hence a
group action ``.'' : $SL(2)\times C^{\infty}(\cM)\to C^{\infty}(\cM)$ of
functions defined as
\beq
(g.f)(z)\~:=\~f(g^{-1}.z)\~, \qquad f \in C^{\infty}(\cM)\~, 
\qquad g\in SL(2)\~, \qquad z\in \cM\~, \label{intrinsicgl2actiononfunctions} 
\eeq
that is compatible
\beq
g.\{f,h\}^{b}\=\{g.f,g.h\}^{a}\~(g^{-1})\q_{a}{}^{b}\~, 
\qquad f,h \in C^{\infty}(\cM)\~, \qquad g\in SL(2)\~, 
\label{compatibleintrinsicgl2action}
\eeq
with the rotations \e{gl2rotationpb} of the two Poisson brackets.
See also Subsection~\ref{secgl2sym}.

\subsection{Bi-Darboux Coordinates}

\noi
General local coordinates are called $z^{A}$,  $A\in\{1,\ldots,3n\}$,
and they are assumed to have definite Grassmann parity 
$\eps_{A}\equiv\eps(z^{A})$. (More precisely, the local coordinates $z^{A}$ are 
functions on an open neighborhood $\cU\subseteq\cM$, and usually not 
globally defined. Nevertheless, we will often, with a slight misuse of 
notation, not explicitly mention the neighborhood $\cU$, and write
$z^{A} \in C^{\infty}(\cM)$, $dz^{A}\in \Gamma(T^{*}\cM)$, etc., rather than
$z^{A} \in C^{\infty}(\cU)$, $dz^{A}\in \Gamma(T^{*}\cM|_{\cU})$, respectively.)

\begin{definition}
{\bf Bi-Darboux coordinates} (or {\bf bi-canonical coordinates}) 
for the two Poisson brackets $\{\cdot,\cdot\}^{a}$, $a\in\{1,2\}$, are a common
set of local Darboux coordinates $\{z^{A}\}=\{q^{i};p\q_{aj}\}$, 
$i,j\in\{1,\ldots,n\}$, $a\in\{1,2\}$, with Grassmann parities 
$\eps_{i}\equiv\eps(q^{i})$ and $\eps(p\q_{aj})=\eps_{j}\!+\!\eps$, such that
\beq
\{f,g\}^{a}= f \left(\papar{q^{i}} \papal{p\q_{ai}}
-(-1)^{\eps_{i} (1-\eps)}\papar{p\q_{ai}}\papal{q^{i}}\right) g\~,
\qquad f,g \in C^{\infty}(\cM)\~,    \qquad a\in\{1,2\}\~. 
\label{bidarbouxform01}
\eeq
\end{definition}

\subsection{Casimirs}

\begin{definition}
A local function $f\in C^{\infty}(\cU)$, $\cU \subseteq \cM$, is by definition a 
{\bf Casimir}\footnote{Casimirs are called {\em marked functions} in
\Ref{grisemi97}, \Ref{grisemi98} and \Ref{gri99}.}
for the $a$'th Poisson bracket $\{\cdot,\cdot\}^{a}$ if the corresponding local 
Hamiltonian vector field $X^{a}_{f}:=\{f,\cdot\}^{a}=0$ vanishes identically. 
\end{definition}

\noi
The subalgebra (more correctly, subsheaf) of Casimirs for the first (second) 
Poisson bracket is denoted $\cC\q_{2}$ ($\cC\q_{1}$), respectively. (Notice the
reversed labeling convention!) The $2n$ rank condition means that the subalgebra
$\cC\q_{a}\subseteq C^{\infty}(\cM)$, $a\in\{1,2\}$, is locally generated by $n$
independent Casimir coordinates $\xi\q_{ai}$, $i\in\{1,\ldots,n\}$. (The
notation $\xi\q_{ai}$ is a bit misleading in the sense that $\xi\q_{ai}$ does 
{\em not} necessarily transform as an $SL(2)$ doublet under $SL(2)$ rotations 
of the ``$a$'' index.) {}For fixed $a\in\{1,2\}$, the set of
local Casimir coordinates $\xi\q_{ai}$ is unique up to reparametrizations
$\xi\q_{ai}\to \xi^{\prime}_{aj}= \xi^{\prime}_{aj}(\xi\q_{a})$.

\noi
The above reversed labeling convention implies that $\{\cdot,\xi\q_{ai}\}^{b}$
is diagonal in the $\q_{a}{}^{b}$ indices. (This choice of labeling convention
is necessary, so that, \eg the formula \e{bidarbouxform01} for bi-Darboux
coordinates becomes manifestly $GL(2)$ covariant under the identification
$p\q_{ai}=\xi\q_{ai}$.)

\noi
The two Poisson brackets $\{\cdot,\cdot\}^{a}$, $a\in\{1,2\}$, are furthermore 
assumed to have the following properties.

\begin{enumerate}
\item
They are {\em jointly non-degenerate}, which means that they have no common 
Casimirs $\cC\q_{1}\cap\cC\q_{2}\subseteq\{0\}$.
\item
They have {\em mutually involutive\footnote{Other names are {\em mutually flat}
or {\em mutually commutative}, \cf \Ref{grisemi97}, \Ref{grisemi98} and
\Ref{gri99}.} Casimirs}, which means that the Casimirs \wrt one bracket are in
involution \wrtt other bracket, and vice-versa. In other words,
\beq
\{f,g\}^{a}\=0\~, \qquad f,g\in\cC\q_{a}\~, \qquad a\in\{1,2\}\~.
\eeq
This can be written compactly as $\{\cC\q_{a},\cC\q_{a}\}^{a}\subseteq\{0\}$;
or equivalently, in local Casimir coordinates,
\beq
 \{\xi\q_{ai},\xi\q_{aj}\}^{a}\=0\~, 
\qquad i,j\in\{1,\ldots,n\}\~, \qquad a\in\{1,2\}\~.\label{mutualflat01}
\eeq
In fact, it follows from \eq{mutualflat01} and the Casimir property, that
$\{\cC\q_{a},\cC\q_{b}\}^{c}\subseteq\{0\}$, or equivalently,
\beq
 \{\xi\q_{ai},\xi\q_{bj}\}^{c}\=0\~, \qquad i,j\in\{1,\ldots,n\}\~, 
\qquad a,b,c\in\{1,2\}\~.\label{mutualflat02}
\eeq
\end{enumerate}

\begin{definition} 
A {\bf triplectic} manifold $(\cM;\{\cdot,\cdot\}^{a})$ is a $3n$-dimensional
manifold $\cM$ equipped with two Poisson brackets $\{\cdot,\cdot\}^{a}$,
$a\in\{1,2\}$, 
\begin{enumerate}
\item
that both have rank $2n$, 
\item
that have common intrinsic Grassmann parity $\eps$,
\item
that are compatible, 
\item
that are jointly non-degenerate, 
\item
and that have mutually involutive Casimirs. 
\end{enumerate}
\label{deftriplectic}
\end{definition}

\subsection{Fiber Bundle $\cM\to\cN$}
\label{secfiberbundle}

\noi
We assume from now on that $(\cM;\{\cdot,\cdot\}^{a})$ is a $3n$-dimensional
triplectic manifold. {}For each Poisson bracket $\{\cdot,\cdot\}^{a}$, 
$a\in\{1,2\}$, there exists an integrable distribution 
$\Delta^{a}=T(\cM\q_{a}) \subseteq T\cM$, generated by the Hamiltonian vector
fields $X^{a}_{f}:=\{f,\cdot\}^{a}$, $f\in C^{\infty}(\cM)$. The distribution 
$\Delta^{a}=T(\cM\q_{a})$ gives rise to a $2n$-foliation of $\cM$ called 
{\em symplectic leaves}. Locally, the $2n$-dimensional symplectic leaves are
labeled by $n$ constants $\xi^{(0)}_{ai}$, $i\in\{1,\ldots,n\}$,
\bea
\left. \cM\q_{1}(\xi^{(0)}_{2i})\right|_{\cU} 
&:=& \{z\in \cU \mid \forall  i\in\{1,\ldots,n\}:\~  
\xi\q_{2i} \= \xi^{(0)}_{2i}\}\~,  \nonumber \\
\left. \cM\q_{2}(\xi^{(0)}_{1i})\right|_{\cU} 
&:=& \{z\in \cU \mid \forall  i\in\{1,\ldots,n\}:\~  
\xi\q_{1i} \= \xi^{(0)}_{1i}\}\~, \label{emmleaf}
\eea
The $n$-dimensional submanifolds 
\beq
\cM\q_{1}(\xi^{(0)}_{2i})\~\cap\~ \cM\q_{2}(\xi^{(0)}_{j1}) \label{qleaf}
\eeq
of intersecting symplectic leaves, are again leaves that constitute an 
$n$-foliation of $\cM$. (The $n$-leaves \e{qleaf} are not necessarily
Lagrangian/involutive, due to possible presence of $F^{aij}$ matrices
\e{abmatrices01}, \cf Section~\ref{secabbakaas}.) 

\noi
Let us collectively call all the $2n$ Casimir coordinates for
$\xi^{I}=\xi\q_{ai}$, where $I\in\{1,\ldots,2n\}$, $i\in\{1,\ldots,n\}$,
$a\in\{1,2\}$. Let the local leaf coordinates (\ie the coordinates that
parametrize a single $n$-leaf) be $q^{i}$, with Grassmann parity
$\eps_{i}\equiv\eps(q^{i})$, $i\in\{1,\ldots,n\}$, in such a way that 
$\{z^{A}\}=\{q^{i};\xi^{I}\}$ constitutes a local coordinate system for
the total space $\cM$. 

\noi
As we shall see in Section~\ref{secposition}, there exists an atlas of 
distinguished\footnote{A {\em distinguished} element of a set means an element 
that has an extra property, which depends on context.}
coordinate systems $\{z^{A}\}=\{q^{i};\xi^{I}\}$ for $\cM$, in-which the
leaf coordinates $q^{i}\to q^{\prime j}$ transform affinely under coordinate
transformations $z^{A}\longrightarrow z^{\prime B}=z^{\prime B}(z)$. 
In other words, an $n$-leaf \e{qleaf} is always (a subsets of) an
$n$-dimensional affine space. 

\noi
{}For this reason, we shall from now on assume the following model for the 
$3n$-dimensional manifold $\cM$ (which locally captures the general
situation).

\begin{assumption}[Fiber bundle]
The triplectic manifold $\cM$ is globally a (not necessarily affine) fiber
bundle $\cM\to \cN$ over a $2n$-dimensional base manifold $\cN$ with local
base coordinates $\xi^{I}$, $I\in\{1,\ldots,2n\}$ consisting of Casimirs. 
(To be more precise, a local Casimir coordinate in $\cM$ is a pull-back
$\pi^{*}\xi\q_{ai}:=\xi\q_{ai}\circ\pi $ of a local coordinate $\xi^{I}$ on
$\cN$ via the canonical projection map $\pi: \cM\to \cN$.) 
\end{assumption}

\noi
The $n$-dimensional fibers have local fiber coordinates 
$q^{i}$, $i\in\{1,\ldots,n\}$.

\subsection{Local Product Manifold $\cN$}
\label{seclocprodmanifold}

\noi
The $2n$-dimensional base manifold $\cN$ has two $n$-foliations with
$n$-dimensional leaves
\bea
\cN\p_{1}(\xi^{(0)}_{2i})\~:=\~\pi(\cM\q_{1}(\xi^{(0)}_{2i}))\~, \qquad
\cN\p_{2}(\xi^{(0)}_{1i})\~:=\~\pi(\cM\q_{2}(\xi^{(0)}_{1i}))\~, \label{ennleaf}
\eea
respectively. Here $\pi: \cM\to \cN$ is the canonical projection map, and here
$\xi^{(0)}_{ai}$ are constants that label the leaves. The $n$-dimensional
tangent space $T(\cN\p_{a})=\pi\q_{*}(\Delta^{a})$ is an integrable distribution
$\subseteq T\cN$. All of this implies that $\cN$ is a {\em local product 
manifold}, which means that there exists an atlas of distinguished coordinate
systems $\{\xi^{I}\}=\{\xi\q_{1i};\xi\q_{2i}\}$ such that a general coordinate
transformation $\xi^{I}\longrightarrow \xi^{\prime J}=\xi^{\prime J}(\xi)$
between two distinguished coordinate systems splits in two sectors,
\beq
\xi\q_{1i}\longrightarrow \xi^{\prime}_{j1}\=\xi^{\prime}_{j1}(\xi\q_{1})\~,
\qquad
\xi\q_{2i}\longrightarrow \xi^{\prime}_{j2}\=\xi^{\prime}_{j2}(\xi\q_{2})\~.
\label{xixisplit}
\eeq

\subsection{$E^{ai}\q_{bj}$ and $F^{aij}$ Matrices}
\label{secabbakaas}

\begin{observation}
In coordinates of the form $\{z^{A}\}=\{q^{i};\xi^{I}\}$, a fundamental
Poisson bracket $\{z^{A},z^{B}\}^{a}$ can only be non-zero if at least
one of the entries $z^{A}$ or $z^{B}$ is a $q^{i}$ variable. 
\end{observation}

\noi
In other words, there are no traces of the bi-Poisson structure on the base
manifold $\cN$ itself, \cf \eq{mutualflat02}. The only remaining non-zero
fundamental Poisson brackets $\{z^{A},z^{B}\}^{a}$ are given by
\beq
E^{ai}\q_{bj}\~:=\~\{q^{i},\xi\q_{bj}\}^{a}\~, 
\qquad F^{aij}\~:=\~ \{q^{i},q^{j}\}^{a}\~, \qquad i,j\in\{1,\ldots,n\}\~, 
\qquad a,b\in\{1,2\}\~. \label{abmatrices01}
\eeq
In fact, one can say more. Note that the $2n\times 2n$ matrix $E^{ai}\q_{bj}$
is diagonal in the ${}^{a}\q_{b}$ indices, due to the Casimir property, and
therefore only consists of two $n\times n$ block matrices, apart from trivial
zero entries. Thus the matrices \e{abmatrices01} effectively only contain four
quadratic $n\times n$ block matrices, where the third and fourth $n\times n$
block matrix come from the $2\times n\times n$ matrix $F^{aij}$. 
The $2n$ rank condition for $\{\cdot,\cdot\}^{a}$ yields the following
Observation~\ref{observationainv}.

\begin{observation}
The two $E^{ai}\q_{aj}$ block matrices are invertible, $a\in\{1,2\}$.
\label{observationainv}
\end{observation}

\noi
\begin{definition}
The $a$'th Poisson bracket $\{\cdot,\cdot\}^{a}$ is said to be on {\bf
Darboux form} (or {\bf canonical form}) if
$E^{ai}\q_{aj}=\delta^{i}_{j}$ and $F^{aij}=0$.
\end{definition}

\section{Bi-Darboux Theorem}

\subsection{Caratheodory-Jacobi-Lie Theorem}
\label{seccajali}

\noi
We now continue dissecting the symmetrized Jacobi identity \e{mixjacid} in a 
triplectic context. To proceed, it is convenient to break the manifest 
$1\leftrightarrow 2$ labeling symmetry between the two Poisson brackets 
$\{\cdot,\cdot\}^{a}$, $a\in\{1,2\}$. We will rename the Casimirs $\xi\q_{ai}$ as
\beq
p\q_{i}\equiv \xi\q_{1i}\~, \qquad c\q_{\jj}\equiv \xi\q_{2\jj}\~, 
\qquad i,\jj\in\{1,\ldots,n\}\~,
\eeq
for reasons that will soon become clear. 

\noi
According to (a superversion of) the Caratheodory-Jacobi-Lie Theorem
\cite{lm87} (with the Casimir $c$ variables as passive spectator parameters),
it is possible to introduce position coordinates $q^{i}$, $i\in\{1,\ldots,n\}$,
such that the first Poisson bracket $\{\cdot,\cdot\}^{1}$ is on Darboux form
\beq
E^{1i}\q_{1j}\~=\~\{q^{i},p\q_{j}\}^{1}\=\delta^{i}_{j}\~, 
\qquad  F^{1ij}\~=\~\{q^{i},q^{j}\}^{1}\=0\~, \qquad i,j\in\{1,\ldots,n\}\~.
\label{darbouxform01}
\eeq
We emphasize that the Darboux form for the first Poisson bracket can be 
achieved {\em without} changing the momenta $p\q_{i}$ and the Casimirs
$c\q_{\jj}$. The Grassmann parity of the momentum variables $p\q_{i}$ must
be $\eps(p\q_{i})=\eps_{i}\!+\!\eps$.

\subsection{$E^{i}\q_{\jj}$ and $F^{ij}$ Matrices}
\label{secabba}

\noi
The only remaining non-zero fundamental brackets $\{z^{A},z^{B}\}^{2}$
for the second Poisson bracket are given by two quadratic $n\times n$ matrices
\beq
E^{i}\q_{\jj}\~:=\~E^{2i}\q_{2\jj}\~:=\~\{q^{i},c\q_{\jj}\}^{2}\~, 
\qquad F^{ij}\~:=\~F^{2ij}\~:=\~ \{q^{i},q^{j}\}^{2}\~, \qquad 
i,j,\jj\in\{1,\ldots,n\}\~. \label{abmatrices02}
\eeq
The Grassmann parities are
$\eps(E^{i}\q_{\jj})=\eps(p\q_{i})\!+\!\eps(c\q_{\jj})$
and $\eps(F^{ij})=\eps_{i}\!+\!\eps\!+\!\eps_{j}$, respectively.

\noi
The second Poisson bracket  $\{\cdot,\cdot\}^{2}$ is on Darboux form if
$E^{i}\q_{\jj}=\delta^{i}_{\jj}$ and $F^{ij}=0$, and in that case we would
have achieved a bi-Darboux form of the two Poisson brackets.

\noi
If one inspects the six-term mixed Jacobi identity \e{mixjacid} in the $qpc$
and $qqp$ sectors, it turns out that five of the six terms vanish because of
\eq{darbouxform01} or the Casimir property. Hence the remaining lone term
must vanish as well, 
\beq 
\{\{q^{i},c\q_{\jj}\}^{2},p\q_{k}\}^{1}\=0\~, \qquad 
\{\{q^{i},q^{j}\}^{2},p\q_{k}\}^{1}\=0\~, \label{efqindep01}
\eeq
respectively. Equation \e{efqindep01} implies that the matrices 
$E^{i}\q_{\jj}\!=\!E^{i}\q_{\jj}(p,c)$ and $F^{ij}\!=F^{ij}(p,c)$ are
independent of the $q$ variables.
This yields the following Observation~\ref{observationnoqdep}.

\begin{observation}
In coordinates $\{z^{A}\}=\{q^{i};\xi^{I}\}$, where the first Poisson
bracket $\{\cdot,\cdot\}^{1}$ is on Darboux form \e{darbouxform01}, the
fundamental Poisson brackets $\{z^{A},z^{B}\}^{a}$ do not depend on
the $q$ variables.
\label{observationnoqdep}
\end{observation}

\noi
In other words, the fundamental Poisson brackets $\{z^{A},z^{B}\}^{a}$ live down
in the base manifold $\cN$.

\subsection{Bi-Darboux Theorem}
\label{secbidarbouxthm}

\noi
We are now ready to state the bi-Darboux Theorem~\ref{theoremfactorization}.

\noi
\begin{theorem}[Bi-Darboux Theorem]
A necessary and sufficient condition for a triplectic manifold 
$(\cM;\{\cdot,\cdot\}^{a})$ to have bi-Darboux coordinates is a 
local {\bf factorization}\footnote{
Theorem~\ref{theoremfactorization} is essentially stated as Theorem 4.3 in
\Ref{grisemi98}. A factorizable $E^{i}\q_{k}$ matrix \e{factorization} is
referred to as a {\em reducible} matrix in \Ref{grisemi97} and \Ref{grisemi98}.
Those papers rely on additional structures (the odd vector fields $V^{a}$),
which is not used here in order to be as general as possible.}
(or {\bf separation of variables}) condition for the $E^{i}\q_{k}$ matrix
\e{abmatrices02}, \ie there should exist matrices $P^{i}\q_{j}=P^{i}\q_{j}(p)$
and $C^{j}\q_{k}=C^{j}\q_{k}(c)$ such that
\beq
E(p,c) \= P(p)\~C(c) \qquad \Leftrightarrow \qquad
E^{i}\q_{k}(p,c) \= P^{i}\q_{j}(p) \~C^{j}\q_{k}(c)\~.
\label{factorization}
\eeq
\label{theoremfactorization}
\end{theorem}

\noi 
We will give a proof of the bi-Darboux Theorem~\ref{theoremfactorization} in
Section~\ref{secproofbidarboux02}.
The factorization \e{factorization} is unique up to a constant invertible 
matrix $K^{i}\q_{j}$, 
\ie
\beq
P \~\longrightarrow \~ PK\~, \qquad C \~\longrightarrow \~ K^{-1}C\~,
\label{uniquesov}
\eeq
because of separation of the $p$ and $c$ variables.
The corresponding differential factorization condition reads
\beq
\papal{p\q_{i}}\left[(\papal{c\q_{\jj}} \Tilde{E}) E\right]\=0
\qquad \Leftrightarrow \qquad
\papal{c\q_{\jj}}\left[  (\papal{p\q_{i}}E)\Tilde{E}\right]\=0\~,
\label{difffactorization}
\eeq
where $\Tilde{E}:=E^{-1}$ denotes the inverse matrix; see also
\eq{defatildematrix}. The differential factorization condition
\e{difffactorization} is equivalent to that the Obata connection $\nabla$
should be flat, see Theorem~\ref{theoremobata}.

\noi
{\em A 3-dimensional example}. Let the triplectic manifold be 
$\cM=\{(q^{1},p\q_{1},c\q_{1})\in \R^3 \mid p\q_{1}+c\q_{1}\neq 0\}$ 
with global coordinates $\{q^{1};p\q_{1};c\q_{1}\}$.
Let the first Poisson bracket $\{\cdot,\cdot\}^{1}$ be on Darboux form, and
let the second Poisson bracket $\{\cdot,\cdot\}^{2}$ have 
$E^{i}\q_{k}$ matrix \e{abmatrices02} equal to
$E^{1}\q_{1}=\{q^{1},c\q_{1}\}^{2}=p\q_{1}+c\q_{1}$. This is a Poisson
pencil \e{mixjacid} that does {\em not} satisfy the factorization condition 
\e{factorization}, and hence {\em no} bi-Darboux coordinates exist.

\section{Closedness Conditions and Poincar\'e Lemma}
\label{secpoincarelemma}

\noi
In this Section~\ref{secpoincarelemma}, we in particular derive the
eqs.\ \e{crucialclosedness}, \es{alphaexact02}{gammaexact02}, which
will be needed in Sections~\ref{secproofbidarboux}--\ref{secproofbidarboux02}. 

\noi
Let $\eta\q_{i}$, $i\in\{1,\ldots,n\}$, be new auxiliary local\footnote{Since we
are only interested in a bi-Darboux Theorem, we may work locally in coordinates.
The word {\em local} refers to a sufficiently small open neighborhood $\cU$.
We will not repeat this point further in the text. Concretely, we will
ignore extending some local constructions to a global setting.} 
``one-form'' variables of Grassmann parity
$\eps(\eta\q_{i})=\eps_{i}\!+\!1\!-\!\eps=\eps(p\q_{i})\!+\!1$.
The Poisson brackets $\{\cdot,\cdot\}^{a}$, $a\in\{1,2\}$, are now trivially
extended such that the $\eta$ variables are new Casimirs for both Poisson 
brackets. Define Grassmann-odd differentials as
\beq
d^{a}\~:=\~\eta\q_{i} \{q^{i},\cdot\}^{a}\= \{\eta\q_{i}q^{i},\cdot\}^{a}\~, 
\qquad \eps(d^{a})\=1\~, \qquad a\in\{1,2\}\~, \label{diffs01}
\eeq  
\beq
d^{1}\=\eta\q_{i}\papal{p\q_{i}}\~,\qquad 
d^{2}\=\eta\q_{i}E^{i}\q_{\jj}\papal{c\q_{\jj}}
+\eta\q_{i}F^{ij}\papal{q^{j}}\~. \label{diffs02}
\eeq
The super-commutator reads
\beq
[d^{a},d^{b}] \= d^{\{a}d^{b\}}
\= \Hf(-1)^{\eps(\eta\q_{j})}\eta\q_{j}\eta\q_{i}
\{\{q^{i},q^{j}\}^{\{a},\cdot\}^{b\}}\=\{\beta^{\{a},\cdot\}^{b\}}\~, 
\qquad a,b\in\{1,2\}\~. 
\label{deltasupercom01}
\eeq
Here 
\beq
\beta^{a}\~:=\~ \Hf(-1)^{\eps(\eta\q_{j})}\eta\q_{j}\eta\q_{i}
\{q^{i},q^{j}\}^{a}\=\Hf\eta\q_{i}
\{q^{i},q^{j}\}^{a}\eta\q_{j}(-1)^{\eps(\eta\q_{j})\eps}\~, 
\qquad a\in\{1,2\}\~, \label{betaaah}
\eeq
are two-forms. The super-commutator \e{deltasupercom01} vanishes if we restrict
the differentials $d^{a}$ to act on an algebra $\cF$ of functions
$f=f(p,c,\eta)$ that do not depend on the $q$ variables. This is basically
because $\{q^{i},q^{j}\}^{a}\in\cF$ does not depend on the $q$'s, \cf 
Observation~\ref{observationnoqdep}. Concretely, the $q^{j}$ differentiation
in \eq{diffs02} becomes irrelevant. The two-forms $\beta^{a}\in\cF$ and the 
one-forms
\beq
\alpha^{a}_{\jj}\~:=\~d^{a}c\q_{\jj}\~\in\~\cF\~, 
 \qquad \jj\in\{1,\ldots,n\}\~, \qquad a\in\{1,2\}~,
\eeq
both belong to $\cF$. It follows from the symmetrized Jacobi identity 
\e{mixjacid} in the $qqq$ and $qqc$ sectors that
\beq
d^{\{a}\beta^{b\}}\=0\~, \qquad a,b\in\{1,2\}\~,\label{mixedclosedness01}
\eeq
and
\beq
d^{\{a}\alpha^{b\}}_{\jj}\=d^{\{a}d^{b\}}c\q_{\jj}\=0\~, 
\qquad a,b\in\{1,2\}\~,\label{mixedclosedness02}
\eeq
respectively. Now, we already know from Section~\ref{seccajali} that the first
structures 
\beq
\beta^{1}\=0 \qquad \mathrm{and} \qquad \alpha^{1}_{\jj}\=0 
\label{betaalpha01}
\eeq
are zero, so we are really only interested in the second structures 
\beq
\beta^{2}\=\Hf\eta\q_{i} F^{ij}\eta\q_{j}(-1)^{\eps(\eta\q_{j})\eps}
\qquad \mathrm{and} \qquad
\alpha^{2}_{\jj}\=\eta\q_{i}E^{i}\q_{\jj}\~.\label{betaalpha02}
\eeq 
It follows that $\beta^{2}$ and $\alpha^{2}_{\jj}$ are $d^{a}$-closed,
\beq
d^{a}\beta^{2}\=0\~, \qquad d^{a}\alpha^{2}_{\jj}\=0\~, \qquad a\in\{1,2\}\~.
\label{crucialclosedness} 
\eeq

\subsection{Closedness Condition for $E^{i}\q_{\jj}$}
\label{secccfora}

\noi
The mixed closedness condition $d^{1}\alpha^{2}_{\jj}=0$ reads explicitly,
\beq
(\papal{p\q_{i}}E^{k}\q_{\jj})\=
(-1)^{\eps(p\q_{i})\eps(p\q_{k})}(i\leftrightarrow k)\~.
\label{alphaclosed02}
\eeq
By the standard Poincar\'e Lemma for $d^{1}$, there exist 
zero-forms $A\q_{\jj}=A\q_{\jj}(p,c)\in\cF$ such that
\beq
\alpha^{2}_{\jj}\=d^{1}A\q_{\jj}\~,\label{alphaexact01}
\eeq
or explicitly,
\beq
E^{i}\q_{\jj}\=(\papal{p\q_{i}}A\q_{\jj})\~.\label{alphaexact02}
\eeq

\subsection{Closedness Condition for $\Tilde{E}^{\jj}\q_{i}$}
\label{secccforta}

\noi
2) The closedness condition $d^{2}\alpha^{2}_{\jj}=(d^{2})^2c\q_{\jj}=0$
reads explicitly,
\beq
E^{i}\q_{\mm} (\papal{c\q_{\mm}}E^{j}\q_{\kk})\=
(-1)^{\eps(p\q_{i})\eps(p\q_{j})} (i\leftrightarrow j)\~,
\label{gammaclosed01}
\eeq
or equivalently,
\beq
\papal{c\q_{\ii}}\Tilde{E}^{\jj}\q_{k}
\=(-1)^{\eps(c\q_{\ii})\eps(c\q_{\jj})}(\ii \leftrightarrow \jj)\~,
\label{gammaclosed02}
\eeq
where we have defined the inverse matrix
\beq
\Tilde{E}^{\jj}\q_{i}\~:=\~\left(E^{-1}\right)^{\jj}\q_{i}\~,
\label{defatildematrix}
\eeq
\cf Observation~\ref{observationainv}.
By the standard Poincar\'e Lemma, there exist functions 
$\Tilde{A}\q_{i}=\Tilde{A}\q_{i}(p,c)\in\cF$, so that
\beq
\Tilde{E}^{\jj}\q_{i}\=(\papal{c\q_{\jj}}\Tilde{A}\q_{i})\~.
\label{gammaexact02}
\eeq

\section{Canonical Transformations and Bundle Structure}
\label{secproofbidarboux}

\subsection{Groupoids $\cG$, $\cG\q_{1}$ and $\cG\q_{2}$}
\label{secgroupoid}

\noi
It is very restricted what local coordinate transformations
$z^{A}\to z^{\prime B}=z^{\prime B}(z)$ can still be performed without spoiling the
progress so far in the attempt to achieve bi-Darboux coordinates.
They are given by the following groupoid $(\cG;\circ)$.

\begin{definition}
Let $(\cG;\circ)$ be the groupoid of local coordinate transformations
$z^{A}\to z^{\prime B}=z^{\prime B}(z)$ that satisfy the 
following conditions.

\begin{itemize}
\item
They preserve the Darboux form \e{darbouxform01} of the first Poisson bracket 
$\{\cdot,\cdot\}^{1}$.
\item 
They at most reparametrize the Casimirs 
$p\q_{i}\to p^{\prime}_{j}= p^{\prime}_{j}(p)$ and
$c\q_{\ii}\to c^{\prime}_{\jj}= c^{\prime}_{\jj}(c)$.
\end{itemize}
\label{defggroupoid}
\end{definition} 

\begin{definition}
Let $\cG\q_{1}\subseteq \cG$ be the subgroupoid of local coordinate 
transformations $z^{A}\to z^{\prime B}=z^{\prime B}(z)$ that do 
{\bf not} transform the second set of Casimirs $c^{\prime}_{\ii}=c\q_{\ii}$.
\label{defggroupoid1}
\end{definition} 

\begin{definition}
Let $\cG\q_{2}\subseteq \cG$ be the subgroupoid of local coordinate 
transformations $z^{A}\to z^{\prime B}=z^{\prime B}(z)$ 
that do {\bf not} transform $q^{\prime i}= q^{i}$ {\bf nor} 
$p^{\prime}_{j}= p\q_{j}$ but do {\bf only} reparametrize the second Casimirs 
$c\q_{\ii}\to c^{\prime}_{\jj}= c^{\prime}_{\jj}(c)$.
\label{defggroupoid2}
\end{definition} 

\noi
The two subgroupoids $\cG\q_{1}$ and $\cG\q_{2}$ commute, and each 
coordinate transformations
$z^{A}\to z^{\prime B}=z^{\prime B}(z)$ in $\cG$ 
may be uniquely factorized in two coordinate transformations
from  $\cG\q_{1}$ and $\cG\q_{2}$, respectively.

\subsection{Canonical Transformations}
\label{secct}

\noi
Let us first consider a coordinate transformation
$z^{A}\to z^{\prime B}=z^{\prime B}(z)$ in just $\cG\q_{1}$,
which preserves the first Poisson bracket on Darboux form and does not 
transform the Casimir $c$ variables. In other words, it is a {\em canonical
transformation} (\wrtt first Poisson bracket and with the Casimir $c$
variables as passive spectator parameters). As mentioned in \Ref{goldstein80}, 
if the canonical transformation is sufficiently close to the identity,
there exists a corresponding generator
$F\p_{3}=F\p_{3}(q^{\prime},p,c)$ of Grassmann parity $\eps(F\p_{3})=\eps$,
which depends on the new positions $q^{\prime i}$ and the old momenta $p\q_{j}$,
such that
\beq
-\dd p\q_{i}\~ q^{i}\= p^{\prime}_{j}\~ \dd q^{\prime j}+\dd F\p_{3}\~,\qquad 
q^{i}\=-(\papal{p\q_{i}}F\p_{3})\~,\qquad 
p^{\prime}_{j}\=-(F\p_{3}\papar{q^{\prime j}})\~.\label{f3transf01}
\eeq
(The most general coordinate transformation in $\cG\q_{1}$ is a finite
composition of $F\p_{3}$ type transformations \e{f3transf01}. This can for
instance be proven \wtho Moser's trick \cite{moser65}.)
The new momenta $p^{\prime}_{j}= p^{\prime}_{j}(p)$ should still be Casimirs
for the second Poisson bracket,
\bea
0&=& \{p^{\prime}_{i}, c\q_{\kk}\}^{2}
\=(p^{\prime}_{i}\papar{q^{\prime j}})\{q^{\prime j},c\q_{\kk}\}^{2}
+(p^{\prime}_{i}\papar{p\q_{j}})\{p\q_{j},c\q_{\kk}\}^{2}
+(p^{\prime}_{i}\papar{c\q_{\jj}})\{c\q_{\jj},c\q_{\kk}\}^{2} \cr
&=& -(F\p_{3}\papar{q^{\prime i}}\papar{q^{\prime j}})E^{\prime j}\q_{\kk}\~.
\label{cascons01}
\eea
Since the new matrix $E^{\prime j}\q_{\kk}$ must be invertible, \cf 
Observation~\ref{observationainv}, the second derivatives of $F\p_{3}$
\wrtt $q^{\prime}$ variables must vanish,
\beq
(F\p_{3}\papar{q^{\prime i}}\papar{q^{\prime j}})\=0\~.
\eeq
Hence the generator 
\beq
-F\p_{3}\=A\q_{j}(p,c)\~q^{\prime j}+B(p,c) \label{f3affine}
\eeq
is affine in the new positions $q^{\prime i}$. 
(The minus sign is introduced for later convenience. At this stage, the 
$F\p_{3}$ coefficient functions $A\q_{j}=A\q_{j}(p,c)$ and $B=B(p,c)$ are
supposed to be independent of any previous definitions.) 
The new momenta $p^{\prime}_{j}$ become the $A\q_{j}$ coefficient functions,
\beq
p^{\prime}_{j}(p)\=-(F\p_{3}\papar{q^{\prime j}})\=A\q_{j}(p,c)\~.
\eeq
In particular, we conclude the following Observation~\ref{observationaindepc}.

\begin{observation}
The $F\p_{3}$ coefficient functions $A\q_{j}=A\q_{j}(p)$ must be independent of 
the $c$ variables.
\label{observationaindepc}
\end{observation}

\subsection{Positions $q^{i}$}
\label{secposition}

\noi
The positions $q^{i}\to q^{\prime j}$ transform affinely under coordinate
transformations $z^{A}\to z^{\prime B}=z^{\prime B}(z)$ in $\cG$,
\beq
q^{i}\=-(\papal{p\q_{i}}F\p_{3})
\= (\papal{p\q_{i}}A\q_{j})q^{\prime j}+(\papal{p\q_{i}}B)\~. 
\label{oldq01}
\eeq
Combined with transformations from $\cG\q_{2}$, \eq{oldq01} proves the
following Proposition~\ref{propqqmatrix}.

\begin{proposition}[Affinity] 
The bundle $\cM \to \cN$ is an affine fiber bundle. Under a coordinate
transformation $z^{A}\to z^{\prime B}=z^{\prime B}(z)$ that
belongs to the groupoid $\cG$, the positions $q^{i}\to q^{\prime j}$ transform
affinely with Jacobian matrix given by 
\beq
(q^{i}\papar{q^{\prime j}}) \=(\papal{p\q_{i}}p^{\prime}_{j}) 
\qquad\Leftrightarrow\qquad
(\papal{q^{\prime j}}q^{i}) 
\= (-1)^{(\eps_{i}+\eps_{j})(1-\eps)}(p^{\prime}_{j}\papar{p\q_{i}})\~.
\eeq
\label{propqqmatrix}
\end{proposition}

\subsection{$E^{i}\q_{\jj}$ matrix}
\label{secaah}

\noi
Returning again to just the $F\p_{3}$ transformation \e{f3transf01} from
Subsection~\ref{secct}, the 
$E^{i}\q_{\kk}$ matrix \e{abmatrices02} transforms
$E^{i}\q_{\kk} \to E^{\prime j}\q_{\kk}$ as a tensor
\bea
E^{i}\q_{\kk} &:=& \{q^{i},c\q_{\kk}\}^{2}
\=(q^{i}\papar{q^{\prime j}})\{q^{\prime j},c\q_{\kk}\}^{2}
+(q^{i}\papar{p\q_{j}})\{p\q_{j},c\q_{\kk}\}^{2}
+(q^{i}\papar{c\q_{\jj}})\{c\q_{\jj},c\q_{\kk}\}^{2} \cr
&=&(\papal{p\q_{i}}A\q_{j})\~E^{\prime j}\q_{\kk}
\=(\papal{p\q_{i}}p^{\prime}_{j})\~E^{\prime j}\q_{\kk}\~.
\label{aijmixedtensor01}
\eea
Combined with transformations from $\cG\q_{2}$, \eq{aijmixedtensor01} proves 
the following Proposition~\ref{propatensor}.

\begin{proposition} 
Under a coordinate transformation
$z^{A}\to z^{\prime B}=z^{\prime B}(z)$ that belongs to the
groupoid $\cG$, the upper (lower) index of the $E^{i}\q_{\jj}$ matrix
\e{abmatrices02} transforms as contravariant (covariant) tensor
\beq
E^{i}\q_{\mm}\=(\papal{p\q_{i}}p^{\prime}_{j})\~
E^{\prime j}\q_{\kk}\~(\papal{c^{\prime}_{\kk}}c\q_{\mm})
\label{atensor}
\eeq
of the corresponding descended coordinate transformation
$p\q_{i}\to p^{\prime}_{j}= p^{\prime}_{j}(p)$
($c\q_{\ii}\to c^{\prime}_{\jj}= c^{\prime}_{\jj}(c)$)
of the local product manifold $\cN$, respectively.
\label{propatensor}
\end{proposition}

\subsection{Para-Dolbeault Differentials}
\label{secparadolbeaultdiff}

\noi
Inspired by the $d^{a}$-differentials \e{diffs01}, we define two sets of 
{\em para-Dolbeault differentials},
\beq
\partial^{a}\~:=\~\dd p\q_{i}\~\{q^{i}, \xi^{I}\}^{a} \papal{\xi^{I}}\~,
\qquad \dd \=\dd z^{A} \papal{z^{A}}
\=\dd q^{i} \papal{q^{i}} + \partial^{1} + \Tilde{\partial}^{1}\~,
\label{defparadolbeault01}
\eeq
\beq
\partial^{1}\~:=\~
\dd p\q_{i} \papal{p\q_{i}}\~, \qquad
\Tilde{\partial}^{1}\~:=\~
\dd c\q_{\jj} \papal{c\q_{\jj}}\~,
\label{defparadolbeault02}
\eeq
\beq
\partial^{2}\~:=\~
\dd p\q_{i}\~E^{i}\q_{\jj}\papal{c\q_{\jj}}\~, \qquad
\Tilde{\partial}^{2}\~:=\~
\dd c\q_{\ii}\~\Tilde{E}^{\ii}\q_{j}\papal{p\q_{j}}\~.
\label{defparadolbeault03}
\eeq
The definitions \e{defparadolbeault01}--\e{defparadolbeault03} are invariant
under local coordinate transformations
$z^{A}\to z^{\prime B}=z^{\prime B}(z)$ in $\cG$, \cf 
Proposition~\ref{propqqmatrix} and Proposition~\ref{propatensor}. 
Note that whereas the $d^{a}$-differentials \e{diffs01} are Grassmann-odd and
have form-degree $0$, the differentials
\e{defparadolbeault01}--\e{defparadolbeault03} are Grassmann-even and have
form-degree $1$, \cf Subsection~\ref{secnotation}.
The $2\times2=4$ para-Dolbeault differentials $\partial^{a}$ and
$\Tilde{\partial}^{b}$ satisfy
\beq
[\partial^{a},\partial^{b}] \= 0\~, \qquad
[\Tilde{\partial}^{a},\Tilde{\partial}^{b}] \= 0\~, \qquad
[\partial^{a},\Tilde{\partial}^{b}] \~\propto\~ \epsilon^{ab}\~,   
\eeq 
because of closedness conditions \es{alphaclosed02}{gammaclosed02}.

\subsection{Presymplectic Potential $\vartheta$}
\label{secsymppot}

\begin{definition}
The subgroupoid $\cG\q_{0} \subseteq \cG$ of {\bf restricted coordinate
transformations} consists of local coordinate transformations
$z^{A}\to z^{\prime B}=z^{\prime B}(z)$ such that the positions 
$q^{i}\to q^{\prime j}$ transform linearly without an
inhomogeneous term.
\label{defggroupoidlin}
\end{definition} 

\begin{definition}
The subgroupoid $\cG\q_{\gauge}\subseteq \cG$ of {\bf gauge transformations}
consists of local coordinate transformations
$z^{A}\to z^{\prime B}=z^{\prime B}(z)$ that do {\bf not}
transform $p^{\prime}_{j}= p\q_{j}$ {\bf nor} $c^{\prime}_{\kk}=c\q_{\kk}$
but do {\bf only} transform the positions
\beq
q^{i}\~\longrightarrow\~ q^{\prime i}\=q^{i}-(\papal{p\q_{i}}B)
\eeq
by an Abelian gauge transformation, where $B\=B(p,c)$ is the gauge parameter.
\label{defggroupoidgauge}
\end{definition} 

\noi
Every coordinate transformations
$z^{A}\to z^{\prime B}=z^{\prime B}(z)$ in $\cG$ 
may be written as a composition of a restricted coordinate 
transformation and a gauge transformation from 
Definitions~\ref{defggroupoidlin}--\ref{defggroupoidgauge}.

\begin{definition}
The {\bf presymplectic potential} $\vartheta$ is defined locally as
\beq
\vartheta\~:=\~\dd z^{A}\~\vartheta\q_{A}\=
\dd q^{i}\~\vartheta\q_{i} + \dd p\q_{j}\~\vartheta^{j} 
+ \dd c\q_{\kk}\~\vartheta^{\kk}
\~:= \~-\dd p\q_{j}\~q^{j}\~, 
\eeq
with components
\beq 
\vartheta\q_{i}\~:=\~0\~, \qquad  \vartheta^{j}\~:=\~-q^{j}\~, \qquad 
\vartheta^{\kk}\~:=\~0\~, \qquad \eps(\vartheta)\=\eps\~.
\eeq
\label{defsymppot}
\end{definition} 

\noi
In other words, the presymplectic potential $\vartheta$ is basically a gadget
to keep track of the fiber coordinates $q^{i}$. The presymplectic potential
$\vartheta$ itself is parallel to the $\cN\p_{1}$ leaves, \ie the restricted
one-form $\vartheta$ has grading $(1,0)$ \wrtt first para-Dolbeault pair
$(\partial^{1},\Tilde{\partial}^{1})$.

\begin{proposition}
The locally defined presymplectic potential $\vartheta$
\begin{enumerate}
\item
behaves as a one-form (=co-vector)
$\vartheta^{i}=(\papal{p\q_{i}}p^{\prime}_{j})\vartheta^{\prime j}$
under restricted coordinate transformations, with bi-grading $(1,0)$ 
\wrtt first para-Dolbeault pair $(\partial^{1},\Tilde{\partial}^{1})$;
\item
and $\vartheta$ behaves as gauge potential 
\beq  
\vartheta\~\longrightarrow\~\vartheta^{\prime}\=\vartheta+(\partial^{1}B)  
\qquad\Leftrightarrow\qquad
\vartheta^{i}\~\longrightarrow\~ 
\vartheta^{\prime i}\=\vartheta^{i}+(\papal{p\q_{i}}B)\~, 
\label{symppotgaugepot}
\eeq
under gauge transformations. 
\end{enumerate}
\label{propositionsymppot}
\end{proposition}

\noi
Proposition~\ref{propositionsymppot} shows that the fiber bundle $\cM\to\cN$ 
has a locally defined gauge potential/connection $\vartheta$, and a globally
defined field strength/curvature 
\beq
\omega\~:=\~\dd \vartheta\=\dd p\q_{i}\wedge \dd q^{i}
\~\in\~\Gamma(\bigwedge{}^{2}(T^{*}\cM))\~,
\eeq
see Subsection~\ref{secgaugebundle}. In particular, the gauge bundle is
{\em never} flat. The presymplectic two-form $\omega$ is invariant under
coordinate transformations
$z^{A}\to z^{\prime B}=z^{\prime B}(z)$ in $\cG$.
It corresponds to the $\cM\q_{1}$ foliations of symplectic leaves 
for the first Poisson bracket $\{\cdot,\cdot\}^{1}$.

\subsection{$F^{ij}$ matrix}
\label{secbeeh}

\noi
Returning one more time to just the $F\p_{3}$ transformation \e{f3transf01} 
from Subsection~\ref{secct}, the $F^{im}$ matrix \e{abmatrices02} transforms
$F^{im} \to F^{\prime jk}$ with an inhomogeneous term
\bea
{}F^{im}
&-&(\papal{p\q_{i}}p^{\prime}_{j})\~F^{\prime jk}\~
(p^{\prime}_{k}\papar{p\q_{m}})(-1)^{(\eps_{k}+\eps_{m})(1-\eps)} \cr
&=&\{q^{i},q^{m}\}^{2} -(q^{i}\papar{q^{\prime j}})\{q^{\prime j},q^{\prime k}\}^{2}
(\papal{q^{\prime k}}q^{m})\cr
&=&(q^{i}\papar{q^{\prime j}})\{q^{\prime j},c\q_{\kk}\}^{2}
(\papal{c\q_{\kk}}q^{m})
+(q^{i}\papar{c\q_{\jj}})\{c\q_{\jj},q^{\prime k}\}^{2}
(\papal{q^{\prime k}}q^{m}) \cr
&=&E^{i}\q_{\kk}\papal{c\q_{\kk}}\{q^{m},B\}^{1}
-(-1)^{\eps(p\q_{i})\eps(p\q_{m})} (i\leftrightarrow m)\cr
&=&\{q^{i},q^{k}\}^{2}\papal{q^{k}}\{q^{m},B\}^{1}
+\{q^{i},p\q_{k}\}^{2}\papal{p\q_{k}}\{q^{m},B\}^{1}\cr
&&+\{q^{i},c\q_{\kk}\}^{2}\papal{c\q_{\kk}}\{q^{m},B\}^{1}
-(-1)^{\eps(p\q_{i})\eps(p\q_{m})} (i\leftrightarrow m)\cr
&=&\{q^{i},\{q^{m},B\}^{1}\}^{2}
-(-1)^{\eps(p\q_{i})\eps(p\q_{m})} (i\leftrightarrow m)\~. \label{bimtransf}
\eea

\begin{proposition}
The locally defined object 
\beq
{}F\~:=\~- \Hf \dd p\q_{j}\wedge \dd p\q_{i} \~F^{ij}
\=\Hf\~ \dd p\q_{i} \~F^{ij}\wedge \dd p\q_{j}(-1)^{\eps_{j}(1-\eps)}\~, \qquad
{}F^{ij}\=F^{ij}(p,c)\~,  \label{ueffen01}
\eeq
which is formed from the $F^{ij}$ matrix \e{abmatrices02}, 
\begin{enumerate}
\item
behaves as a two-form 
$F^{im}\=(\papal{p\q_{i}}p^{\prime}_{j})\~F^{\prime jk}\~
(p^{\prime}_{k}\papar{p\q_{m}})(-1)^{(\eps_{k}+\eps_{m})(1-\eps)}$
under restricted coordinate transformations, with bi-grading $(2,0)$ 
\wrtt first para-Dolbeault pair $(\partial^{1},\Tilde{\partial}^{1})$;
\item
and $F$ behaves as    
\bea
{}F&\longrightarrow&F^{\prime}\=F - (\partial^{2}\partial^{1}B)
\=F + (\partial^{1}\partial^{2}B) 
\label{betagauge} \\ &\Updownarrow& \nonumber \\ 
{}F^{ij}&\longrightarrow& F^{\prime ij} 
\= F^{ij}- \left(E^{i}\q_{\kk}(\papal{c\q_{\kk}}\papal{p\q_{j}}B)
-(-1)^{\eps(p\q_{i})\eps(p\q_{j})} (i\leftrightarrow j) \right)\~, 
\label{bijgauge} 
\eea
under gauge transformations.
\end{enumerate}
\label{propositionbeta}
\end{proposition}

\noi
The restricted two-form $F$ from \eq{ueffen01} corresponds to the
``two-form'' $\beta^{2}$ from \eq{betaaah}. The $d^{a}$-closedness condition
\e{crucialclosedness} for the two-form $\beta^{2}$ translates into that the
two-form $F$ is $\partial^{a}$-closed,
\beq
(\partial^{a}F)\=0\~.
\eeq

\subsection{Gauge Bundle}
\label{secgaugebundle}

\noi
We now rephrase the fiber bundle construction using the language of gauge
bundles. 

\begin{enumerate}
\item
{}From the perspective of a gauge bundle over $\cN$, the groupoid $\cG\q_{0}$
of restricted coordinate transformations become by definition the only allowed 
coordinate transformations. Then the fiber bundle $\cM\to \cN$ becomes a linear
vector bundle; and $\vartheta\q_{(\alpha)}\in
\Gamma(\left.T^{*}\cM\right|\q_{\R^{n}\times\cU\q_{(\alpha)}})$ and
$F\p_{(\alpha)}\in
\Gamma(\left.\bigwedge{}^{2}(T^{*}\cN)\right|\q_{\cU\q_{(\alpha)}})$ become two
families of differential forms, which are labeled by local neighborhoods
$\cU\q_{(\alpha)}\subseteq\cN$. 
\item
It should be stressed that the word {\em gauge bundle} in this paper is used
in a slightly non-standard way. Although $\vartheta\q_{(\alpha)}$ plays the
r\^{o}le of a gauge potential/connection, it is not an ordinary gauge
potential, since besides dependence on the base coordinates
$\xi^{I}_{(\alpha)}$, it also depends on the fiber coordinates
$q^{i}_{(\alpha)}$. Another peculiarity is that a change of the base coordinates
$p\q_{(\alpha)i}\to p^{\prime}_{(\alpha)j}= p^{\prime}_{(\alpha)j}(p_{(\alpha)j})$
induces a corresponding change in the fiber coordinates 
$q^{i}_{(\alpha)}
=(\papal{p\q_{(\alpha)i}}p^{\prime}_{(\alpha)j})q^{\prime j}_{(\alpha)}$, \cf
Proposition~\ref{propqqmatrix}.
\item
A gauge transformation from an $(\alpha)$-gauge in a local patch
$\cU\q_{(\alpha)}$ to a $(\beta)$-gauge in a local patch $\cU\q_{(\beta)}$ makes 
sense if the overlap $\cU\q_{(\alpha)}\cap\cU\q_{(\beta)}\neq \emptyset$ is
non-empty. The gauge transformation is  
\beq
q^{i}_{(\beta)}\=q^{i}_{(\alpha)}-(\papal{p\q_{i}}B\q_{(\alpha\beta)})\~, \qquad
\vartheta\q_{(\beta)}\=\vartheta\q_{(\alpha)}+(\partial^{1}B\q_{(\alpha\beta)})\~,
\qquad
{}F\p_{(\beta)}\=F\p_{(\alpha)} +(\partial^{1}\partial^{2}B\q_{(\alpha\beta)})\~, 
\eeq
with gauge parameter $B\q_{(\alpha\beta)}=B\q_{(\alpha\beta)}(\xi)$.
\item
{}For a triple overlap 
$\cU\q_{(\alpha)}\cap\cU\q_{(\beta)}\cap\cU\q_{(\gamma)}\neq\emptyset$,
one must demand the cocycle condition
\beq
B\q_{(\alpha\beta)}+B\q_{(\beta\gamma)}+B\q_{(\gamma\alpha)}
\=C\q_{(\alpha\beta\gamma)} 
\eeq  
for some functions $C\q_{(\alpha\beta\gamma)}$ with 
$(\partial^{1}C\q_{(\alpha\beta\gamma)})=0$, \ie functions 
$C\q_{(\alpha\beta\gamma)}=C\q_{(\alpha\beta\gamma)}(c)$ that only depend
on the $c$ coordinates. 
\end{enumerate}

\section{Proof of Bi-Darboux Theorem}
\label{secproofbidarboux02}

\subsection{Factorization Condition}

\noi
We next continue with the proof of bi-Darboux
Theorem~\ref{theoremfactorization}. Note that
Proposition~\ref{propatensor} shows immediately that the factorization
condition \e{factorization} is necessary for the bi-Darboux
Theorem~\ref{theoremfactorization}. 

\noi
On the other hand, let us from now on assume that the $E=PC$ factorization
condition \e{factorization} is satisfied. It then follows from the two
closedness conditions \es{alphaclosed02}{gammaclosed02} that the $P$
and $C$ matrix factors are Jacobi matrices, \ie there exist locally some
reparametrizations $p\q_{i}\to p^{\prime}_{j}= p^{\prime}_{j}(p)$ and
$c\q_{\ii}\to c^{\prime}_{\jj}= c^{\prime}_{\jj}(c)$, such that 
\beq
P^{i}\q_{j}\=(\papal{p\q_{i}}p^{\prime}_{j})\~, \qquad 
\left(C^{-1}\right)^{\ii}\q_{\jj}\= (\papal{c\q_{\ii}}c^{\prime}_{\jj})\~,
\label{pcjac}
\eeq
respectively. Thus by choosing the $F\p_{3}$ coefficient functions $A\q_{j}$ in
\eq{f3affine} to be the new $p^{\prime}_{j}$ variables \e{pcjac} (and letting
the $B$ function in \eq{f3affine} be arbitrary, \eg zero), it is possible to
perform a $F\p_{3}$ transformation in $\cG\q_{1}$, and a reparametrization
$c\q_{\ii}\to c^{\prime}_{\jj}= c^{\prime}_{\jj}(c)$ in $\cG\q_{2}$, such
that the new $E^{\prime j}\q_{\kk}$ matrix \e{abmatrices02} becomes the unit
matrix $E^{\prime j}\q_{\kk}=\delta^{j}_{\kk}$.

\noi
It still remains to show that the new $F^{\prime jk}$ matrix \e{abmatrices02}
can be chosen to be zero $F^{\prime jk}=0$. This will be done in the next
Subsection~\ref{secbipoincarelemma} \wtho 
bi-Poincar\'e Lemma~\ref{lemmabipoincarelemma}.

\subsection{Bi-Poincar\'e Lemma}
\label{secbipoincarelemma}

\noi
Let us from now on assume that the $E^{i}\q_{\jj}$ matrix \e{abmatrices02} is
the unit matrix $E^{i}\q_{\jj}=\delta^{i}_{\jj}$. Now recall the two $d^{a}$
differentials \e{diffs02} and the $d^{a}$-closedness condition
\e{crucialclosedness} for the two-form $\beta^{2}$ in
Section~\ref{secpoincarelemma}. Treating the $q^{i}$ variables as passive
spectator parameters, we are now in the position to apply the
bi-Poincar\'e Lemma~\ref{lemmabipoincarelemma} with the triple
$\{p\q_{i};c\q_{\jj};\eta\q_{k}\}$ as active variables
$\{x^{i}_{1};x^{\jj}_{2};x^{k}_{3}\}$. There hence exists a zero-form
$B=B(p,c)\in\cF$, of Grassmann parity $\eps(B)=\eps$, such that
\beq
\beta^{2}\=d^{2}d^{1}B \=\eta\q_{i}
\{q^{i},\{q^{j},B\}^{1}\}^{2}\eta\q_{j}(-1)^{\eps(\eta\q_{j})\eps}\~,
\label{betaexact01}
\eeq
or explicitly,
\beq
{}F^{ij}\~:=\~\{q^{i},q^{j}\}^{2}\= \{q^{i},\{q^{j},B\}^{1}\}^{2}
 - (-1)^{(\eps_{i}+\eps)(\eps_{j}+\eps)} (i\leftrightarrow j)\~.
\label{betaexact02}
\eeq
By shifting the $q$ variables as
\beq
q^{i} \quad \longrightarrow  \quad 
q^{\prime i} \= q^{i} - \{q^{i},B\}^{1}\= q^{i} - (\papal{p\q_{i}}B)\~,
\eeq
we achieve that the matrix
\beq 
{}F^{ij}\~:=\~ \{q^{i},q^{j}\}^{2}
\quad \longrightarrow  \quad 
{}F^{\prime ij}\~:=\~\{q^{\prime i},q^{\prime j}\}^{2}\=0
\eeq
vanishes, while all the other fundamental Poisson brackets
$\{z^{A},z^{B}\}^{a}$ remain unchanged. 
Or equivalently, we may note by comparing \eqs{bimtransf}{betaexact02} that
the canonical transformation 
\beq
-F\p_{3}\=p\q_{j}\~q^{\prime j} + B
\eeq
leads to $F^{\prime jk}=0$. 
We have thus achieved a canonical form for the second Poisson bracket,
and thereby confirmed that the factorization condition \e{factorization} is
sufficient for the bi-Darboux Theorem~\ref{theoremfactorization}.  
\proofbox

\section{Bi-Canonical Transformations}
\label{secbict}

\noi
Let there be given a $3n$-dimensional triplectic manifold
$(\cM;\{\cdot,\cdot\}^{a})$. 

\begin{definition}
A {\bf bi-canonical transformation} is a coordinate transformation 
$\{z^{A}\}=\{q^{i}; p\q_{aj}\} \longrightarrow
\{z^{\prime A}\}=\{q^{\prime i}; p^{\prime}_{aj}\}$
between two bi-Darboux coordinate systems \e{bidarbouxform01} of positions 
and momenta. 
\label{defbict}
\end{definition}

\begin{proposition}
Necessary conditions for a bi-canonical transformation 
$z^{A}\to z^{\prime B}=z^{\prime B}(z)$ are the following.
\begin{enumerate}
\item
The momenta $p\q_{ai}$ (\ie the Casimirs $\xi\q_{ai}$) transform under rigid
affine reparametrizations $p\q_{ai}\to p^{\prime}_{aj}= p^{\prime}_{aj}(p\q_{a})$
for each $a\in\{1,2\}$, with common constant $n\times n$ Jacobi matrix 
\beq
J^{i}\q_{j}\~:=\~
(\papal{p\q_{ai}}p^{\prime}_{aj})\~, 
\label{jacmat01}
\eeq
(no sum over $a$). In particular, the Jacobi matrix $J^{i}\q_{j}$ must be
independent of $a\in\{1,2\}$.
\item
The transformation of the position coordinates 
$q^{i}\=J^{i}\q_{j} q^{\prime j}+b^{i}$ is composed of a rigid 
constant rotation with the Jacobi matrix \e{jacmat01}
plus a shift $b^{i}=b^{i}(p)$. 
\end{enumerate}
\end{proposition}

\noi
Given a bi-canonical transformation $z\to z^{\prime}$, one can locally
always perform an additional restricted bi-canonical transformation,
$z^{\prime}\to z^{\prime\prime}$
\beq
q^{\prime\prime i}\=J^{i}\q_{j} q^{\prime j}\~, \qquad 
p^{\prime}_{aj}\=p^{\prime\prime}_{ai} J^{i}\q_{j}\~,
\eeq
involving the same constant Jacobi matrix \e{jacmat01}, 
so that the combined bi-canonical transformation
$z\to z^{\prime}\to z^{\prime\prime}$ is just a gauge transformation, \cf
Definitions~\ref{defggroupoidlin}--\ref{defggroupoidgauge}. The following 
Proposition~\ref{propositionbidarbouxgauge} is a consequence of
Proposition~\ref{propositionbeta}.

\begin{proposition}
A necessary and sufficient condition for a bi-canonical gauge transformation 
$q^{i}\to q^{\prime i}\=q^{i}-b^{i}$ is that locally the shift $b^{i}=b^{i}(p)$
is a gradient \wrt both sets of
momenta, \ie there locally exist $B^{a}=B^{a}(p)$, $a\in\{1,2\}$, such that 
\beq
b^{i}\=(\papal{p\q_{ai}}B^{a})
\eeq
(no sum over $a$).
\label{propositionbidarbouxgauge}
\end{proposition}

\noi
The main lesson is that bi-canonical transformations are rigid, in contrast
to standard canonical transformations, which figuratively speaking, exhibit 
flexible behavior, which is capable of washing out local features.

\section{Para-Hypercomplex Structure}
\label{secparahypercomplex}

\subsection{Almost Parity Structures}
\label{secalmostsigmapee}

\noi
An {\em almost parity structure} $P:\Gamma(T\cN)\to \Gamma(T\cN)$,
\beq
P \= \partial^{r}_{I}\~P^{I}\q_{J}\~\otimes\~\larrow{\dd\xi^{J}}\~, 
\qquad  
\eps(P^{I}\q_{J})\=\eps_{I}+\eps_{J}\~,\label{pee}
\eeq
(\aka an {\em almost para-complex structure} or an {\em almost local product
structure}) is a (mixed contravariant and covariant) tensor $P^{I}\q_{J}$
that satisfies \cite{b97}
\beq
P^{2}\=\Id\=\partial^{r}_{I}\~\otimes\~\larrow{\dd \xi^{I}} \~, 
\qquad \eps(P)\=0\~, \qquad 
{\rm dim~ker}(P\pm\Id)\=n\~\equiv\~\Hf{\rm dim}(\cN) \~.\label{peepee}
\eeq
Here $\partial^{r}_{I}\!\equiv\!(-1)^{\eps_{I}}\partial^{\ell}_{I}$ are not
usual partial derivatives. In particular, they do not act on the tensor 
$P^{I}\q_{J}$ in \eq{pee}. Rather they are a
dual basis to the one-forms $\larrow{\dd x^{I}}$:
\beq
\larrow{\dd x^{I}}(\partial^{r}_{J})\=\delta^{I}_{J}\~.\label{dualbases}
\eeq
Phrased differently, the $\partial^{r}_{I}$ are merely bookkeeping devices,
that transform as right partial derivatives under general coordinate
transformations. (To be able to distinguish them from true partial derivatives,
the differentiation variable $\xi^{I}$ on a true partial derivative 
$\partial/\partial \xi^{I}$ is written explicitly.)
It is convenient to introduce two idempotent projection operators
\beq
P\p_{\pm}\~:=\~\Hf(\Id \pm P)\~, \qquad \Id\=P\p_{+}+P\p_{-}\~, \qquad 
P\=P\p_{+}-P\p_{-}\~,\qquad P\p_{\pm}P\p_{\pm}\=P\p_{\pm}\~, \qquad  
P\p_{\pm}P\p_{\mp}\=0\~.\label{peepm}
\eeq

\subsection{Parity Structures}
\label{secintsigmapee}

\noi
We start by defining two {\em chiral Nijenhuis tensors}
$N_{\pm}\~:\~\Gamma(T\cN) \times \Gamma(T\cN) \to \Gamma(T\cN)$,
\beq
 N\q_{\pm}(X,Y)\~:=\~P\p_{\mp}[P\p_{\pm}X,P\p_{\pm}Y]
 \=-(-1)^{\eps(X)\eps(Y)}N\q_{\pm}(Y,X)\~,
\eeq 
where $X,Y \in \Gamma(T\cN)$ are vector fields.
Note that 
\beq 
N\q_{\pm}(X,PY)\=\pm N\q_{\pm}(X,Y)\=N\q_{\pm}(PX,Y)\~.
\eeq
The {\em Nijenhuis tensor}
$N\=\Hf \partial^{r}_{K}\~N^{K}\q_{IJ}\~\otimes\~
\larrow{\dd \xi^{J}} \wedge  \larrow{\dd \xi^{I}}\~
\in~\Gamma\left(T\cN \otimes \bigwedge{}^{2}(T^{*}\cN)\right)$ 
is defined as
\beq
 N\~:=\~4(N\q_{+}+N\q_{-})\~,
\eeq
or 
\beq
N(X,Y)\=[X,Y]+[PX,PY]-P[X,PY]-P[PX,Y]\=-(-1)^{\eps(X)\eps(Y)}N(Y,X)\~.
\eeq
Equivalently in components,
\beq 
-\larrow{\dd \xi^{K}}(N(\partial^{r}_{I},\partial^{r}_{J}))
\= N^{K}\q_{IJ} \= \left(
(P^{K}\q_{I}\papar{\xi^{M}})\~P^{M}\q_{J}
-P^{K}\q_{M}\~(P^{M}\q_{I}\papar{\xi^{J}}) \right)
-(-1)^{\eps_{I}\eps_{J}}(I \leftrightarrow J)\~.
\eeq 
The relation can be inverted to give
\bea
 8 N\q_{\pm}(X,Y)&=&N(X,Y) \pm N(X,P Y)\=2 N(X,P\p_{\pm}Y) \~, \\
8 N^{K}_{\pm}\q_{IJ}
&=&N^{K}\q_{IM}\~P_{\pm}^{M}\q_{J}
-(-1)^{\eps_{I}\eps_{J}}(I \leftrightarrow J)  \~.
\eea

\begin{definition} 
An almost parity structure $P:\Gamma(T\cN)\to\Gamma(T\cN)$ becomes a {\bf
parity structure} if the two chiral Nijenhuis tensors $N\q_{\pm}=0$ vanish.
\end{definition}

\noi
One may show that the two chiral Nijenhuis tensors $N\q_{\pm}=0$ vanish iff the
corresponding Nijenhuis tensor $N=0$ vanishes. The existence of a parity
structure $P:\Gamma(T\cN)\to\Gamma(T\cN)$ implies that
$P\p_{\pm}(T\cN)\subseteq T\cN$ are two integrable distributions, and that the
holonomy of the manifold $\cN$ is $\subseteq GL(n) \times GL(n)$.

\subsubsection{Parity Structure $\Sigma$}
\label{secsigma}

\noi
Recall from Section~\ref{seclocprodmanifold} that the base manifold $\cN$ is a
local product manifold with local coordinates   
\beq
\xi^{I}\=\twobyone{p\q_{i}}{c\q_{\ii}}\~,
\qquad \partial^{\ell}_{I}
\=\twobyone{\partial^{i}_{\ell}}{\partial^{\ii}_{\ell}}\~.
\eeq
An obvious choice of parity structure, which we will call
$\Sigma: \Gamma(T\cN)\to\Gamma(T\cN)$, preserves (inverts) all the tangent
directions $\subseteq T\cN$ of the $\cN\p_{1}$ leaves ($\cN\p_{2}$ leaves)
\e{ennleaf}, respectively,
\beq
\Sigma\~:=\~\Sigma\q_{+}-\Sigma\q_{-}\~, \qquad 
\Sigma\q_{+}\~:=\~\partial^{i}_{r}\~\otimes\~\larrow{\dd p\q_{i}}\~, \qquad 
\Sigma\q_{-}\~:=\~\partial^{\ii}_{r}\~\otimes\~\larrow{\dd c\q_{\ii}} \~,
\eeq
\beq
\Sigma^{I}\q_{J}
\~:=\~\twobytwo{\delta^{j}_{i}}{0}{0}{-\delta^{\jj}_{\ii}}\~, \qquad
\Sigma(\partial^{i}_{\ell})\~:=\~\partial^{i}_{\ell}\~, \qquad
\Sigma(\partial^{\jj}_{\ell})\~:=\~-\partial^{\jj}_{\ell} \~.
\eeq 
The matrix $\Sigma^{I}\q_{J}$ behaves a mixed tensor under coordinate 
transformations of $\cN$. The first pair \e{defparadolbeault02} of 
para-Dolbeault differentials satisfies
\beq
\partial^{1}\=\Sigma^{T}_{+}\dd \~, \qquad 
\Tilde{\partial}^{1}\=\Sigma^{T}_{-}\dd\~, \qquad
\Sigma\p_{\pm}\~:=\~\Hf(\Id \pm \Sigma) \~.
\eeq
Here $\Sigma^{T}_{+}:\Gamma(T^{*}\cN)\to\Gamma(T^{*}\cN)$ and
$\Sigma^{T}_{-}:\Gamma(T^{*}\cN)\to\Gamma(T^{*}\cN)$ are the projection
operators to the $\cN\p_{1}$ and $\cN\p_{2}$ leaf directions, respectively; see
also \eqs{peepm}{transposedformulation01}.

\subsubsection{Parity Structure $P$}
\label{secpee}

\noi
The invertible $E^{i}\q_{\jj}$ matrix \e{abmatrices02} yields another parity
structure
\beq
P\~:=\~
\partial^{i}_{r}\~\Tilde{E}\q_{i}{}^{\jj}\~\otimes\~\larrow{\dd c\q_{\jj}}
\~+\~\partial^{\ii}_{r}\~E\q_{\ii}{}^{j}\~\otimes\~\larrow{\dd p\q_{j}} \~,
\eeq
\beq
P^{I}\q_{J}
\~:=\~\twobytwo{0}{\Tilde{E}\q_{i}{}^{\jj}}{E\q_{\ii}{}^{j}}{0}\~, \qquad
P(\partial^{i}_{\ell})
\~:=\~E^{i}\q_{\jj}\~\partial^{\jj}_{\ell}\~, \qquad
P(\partial^{\jj}_{\ell})
\~:=\~\Tilde{E}^{j}\q_{i}\~\partial^{i}_{\ell} \~,
\eeq 
where we have defined transposed matrices
\beq
P\q_{I}{}^{J}\~:=\~\~(-1)^{\eps_{I}(\eps_{J}+1)}P^{J}\q_{I}\~, 
\qquad 
E\q_{\jj}{}^{i}\~:=\~(-1)^{\eps(p\q_{i})(\eps(c\q_{\jj})+1)}E^{i}\q_{\jj}\~,
\qquad 
\Tilde{E}\q_{i}{}^{\jj}
\~:=\~(-1)^{\eps(c\q_{\jj})(\eps(p\q_{i})+1)}\Tilde{E}^{\jj}\q_{i}\~.
\label{transposedapmatrices}
\eeq
There is an equivalent transposed formulation
$P^{T}:\Gamma(T^{*}\cN)\to\Gamma(T^{*}\cN)$ on the cotangent space,
\beq
P^{T} \= \dd \xi^{I}\~P\q_{I}{}^{J}\~\otimes\~\larrow{i^{\ell}_{J}} 
\= \dd p\q_{i}\~E^{i}\q_{\jj}\~\otimes\~\larrow{i^{\jj}_{\ell}}
\~+\~ \dd c\q_{\ii}\~\Tilde{E}^{\ii}\q_{j}\~\otimes\~\larrow{i^{j}_{\ell}}\~,
\qquad \larrow{i^{\ell}_{J}}(\dd \xi^{I})\=\delta^{I}_{J}\~, 
\label{transposedformulation01}
\eeq
\beq
P\q_{I}{}^{J}\=\twobytwo{0}{E^{i}\q_{\jj}}{\Tilde{E}^{\ii}\q_{j}}{0}\~,
\qquad
P^{T}(\dd c\q_{\jj})\= \dd p\q_{i}\~E^{i}\q_{\jj}\~, \qquad
P^{T}(\dd p\q_{i})\=\dd c\q_{\jj}\~\Tilde{E}^{\jj}\q_{i}\~.
\eeq
The identity $P^{2}=\Id$ follows because $\Tilde{E}:=E^{-1}$ is the
inverse of the $E$ matrix \e{abmatrices02}.
The vanishing of the corresponding Nijenhuis tensor $N=0$ follows from the
integrability conditions \eqs{alphaclosed02}{gammaclosed02}. The second pair
\e{defparadolbeault03} of para-Dolbeault differentials satisfies
\beq
\partial^{2}\=P^{T}\Tilde{\partial}^{1}\~, \qquad 
\Tilde{\partial}^{2}\=P^{T}\partial^{1}\~.
\eeq

\subsection{Para-Hypercomplex Structure}
\label{secso21}

\noi
The two parity structures $\Sigma$ and $P$ from 
Subsections~\ref{secsigma}--\ref{secpee} anticommute
\beq
\{\Sigma,P\}\q_{+} \~:=\~\Sigma P + P\Sigma \=0\~. \label{sigmapeeanticom} 
\eeq
Conversely, any parity structure may be locally diagonalized to the form of
$\Sigma:T\cN\to T\cN$ given in Subsection~\ref{secsigma}. This is just
rephrasing the fact that a manifold equipped with a parity structure is the
same as a local product manifold, \cf Subsection~\ref{seclocprodmanifold}.
Moreover, it is easy to see that any second parity structure $P:T\cN\to T\cN$ 
that anticommute \eq{sigmapeeanticom} must then be of the form given in
Subsection~\ref{secpee} for {\em some} matrix $E^{i}\q_{\jj}$
that satisfies integrability conditions \es{alphaclosed02}{gammaclosed02}. 

\noi
We may then define a complex structure as 
\beq
J\~:=\~P\Sigma\~, \qquad J^{2}\=-\Id\~.
\eeq
Together $\{\Sigma; P; J \}$ span a {\em para-hypercomplex structure}. See also
Subsection~\ref{secgl2sym}.

\begin{theorem}
A triplectic fiber bundle $(\cM\to\cN;\{\cdot,\cdot\}^{a})$ is a
para-hypercomplex gauge bundle with a $\partial^{a}$-closed $(2,0)$-form $F$. 
Conversely, for a given para-hypercomplex gauge bundle $\cM\to\cN$ with a
$\partial^{a}$-closed $(2,0)$-form $F$, the total space $\cM$ may be endowed
with a triplectic structure $\{\cdot,\cdot\}^{a}$, $a\in\{1,2\}$.
\label{theoremonetoone} 
\end{theorem}

\noi
Here the $(2,0)$ bi-grading refers to the first para-Dolbeault pair 
$(\partial^{1},\Tilde{\partial}^{1})$.
The one-to-one correspondence in Theorem~\ref{theoremonetoone} holds, 
basically because all possible consequences of the symmetrized Jacobi identity
\e{mixjacid} have been completely transcribed into the gauge bundle language, 
\cf Subsection~\ref{secgaugebundle}. 
Note that the dimension of a para-hypercomplex manifold $\cN$ must be a
multiplum of $2$ (unlike a hypercomplex manifold, whose dimension must always
be a multiplum of $4$.) 
 
\noi
{\em A 2-dimensional example}. Let the manifold be $\cN=\C=\R^{2}$
with global coordinates $\{\xi^{I}\}=\{p\q_{1};c\q_{1}\}$. Let 
\beq
\Sigma\=\twobytwo{1}{0}{0}{-1}\~, \qquad 
P\=\twobytwo{0}{1}{1}{0}\~, \qquad 
J\=\twobytwo{0}{-1}{1}{0}\~. 
\label{sigmasexample}
\eeq
Let the non-zero fundamental Poisson brackets be
$\{q^{1},p\q_{1}\}^{1}=1=\{q^{1},c\q_{1}\}^{2}$.

\subsection{The Obata Connection $\nabla$}
\label{secobata}

\begin{proposition}[Superversion of the Obata connection \cite{obata56}]
There exists a unique torsion-free connection
$\nabla: \Gamma(T\cN)\times\Gamma(T\cN) \to \Gamma(T\cN)$, that preserves 
the two anticommuting parity structures $\Sigma$ and $P$, \ie
\beq
\nabla\Sigma\=0\~,\qquad \nabla P\=0\~. \label{sigmapeecovperserv01}
\eeq
\end{proposition}

\noi
{\sc Proof}:\~\~
The second condition in \eq{sigmapeecovperserv01} reads in components
\beq
0\=(\nabla^{\ell}_{I} P)^{J}\q_{K}
\= (\papal{\xi^{I}} P^{J}\q_{K})
+\Gamma\q_{I}{}^{J}\q_{M}\~P^{M}\q_{K}
-(-1)^{\eps_{I}\eps_{J}}
P^{J}\q_{M}\Gamma^{M}\q_{IK}\~, \label{sigmapeecovperserv02}
\eeq
where by definition
\beq
\Gamma\q_{I}{}^{J}\q_{K}
\=(-1)^{\eps_{I}\eps_{J}}\Gamma^{J}\q_{IK}\~.
\eeq
The torsion-free condition $T=0$ means that the Christoffel symbols are
graded symmetric in the lower indices
\beq
\Gamma^{K}\q_{IJ}
\=-(-1)^{(\eps_{I}+1)(\eps_{J}+1)}\Gamma^{K}\q_{JI}\~.
\eeq
We may take $\Sigma$ and $P$ as in Subsections~\ref{secsigma}--\ref{secpee}. 
The first condition in \eq{sigmapeecovperserv01} shows that all the mixed
components of the Christoffel symbols $\Gamma^{K}\q_{IJ}$ vanish.
The remaining two non-mixed components can be deduced of from 
\eq{sigmapeecovperserv02}. 
\beq
-\Gamma^{i}\q_{j}{}^{k}
\=(\papal{p\q_{i}}\Tilde{E}\q_{j}{}^{\mm}) E\q_{\mm}{}^{k}
\=E^{i}\q_{\mm}(\Tilde{E}^{\mm}\q_{j}\papar{p\q_{k}})\~, \qquad
(-1)^{\eps(p\q_{i})}\Gamma\q_{k}{}^{ij}
\=\Tilde{E}\q_{k}{}^{\mm}(E\q_{\mm}{}^{i}\papar{p\q_{j}})\~,
\eeq
\beq
-\Gamma^{\ii}\q_{\jj}{}^{\kk}
\=(\papal{c\q_{\ii}}E\q_{\jj}{}^{m})\Tilde{E}\q_{m}{}^{\kk}
\=\Tilde{E}^{\ii}\q_{m}(E^{m}\q_{\jj}\papar{c\q_{\kk}})\~,
\qquad
(-1)^{\eps(c\q_{\ii})}\Gamma\q_{\kk}{}^{\ii\jj}
\=E\q_{\kk}{}^{m}(\Tilde{E}\q_{m}{}^{\ii}\papar{c\q_{\jj}})\~.
\eeq
\proofbox

\begin{theorem}
The Obata connection $\nabla$ is flat if and only if the factorization
condition \e{factorization} holds\footnote{Theorem~\ref{theoremobata} is
essentially stated as Proposition 4.2 in \Ref{grisemi98}}.
\label{theoremobata}
\end{theorem}

\noi
In other words, a triplectic fiber bundle $(\cM\to\cN;\{\cdot,\cdot\}^{a})$
has locally bi-Darboux coordinates iff the Obata connection $\nabla$ on the
para-hypercomplex manifold $\cN$ is flat.

\subsection{Global $GL(2)$ Covariance}
\label{secgl2sym}

\noi
Recall that the $gl(2)$ Lie algebra generators 
\beq
t\q_{0}\=\twobytwo{1}{0}{0}{1}\~, \qquad 
t\q_{1}\=\twobytwo{0}{1}{1}{0}\~, \qquad 
t\q_{2}\=\twobytwo{0}{-1}{1}{0}\~,\qquad  
t\q_{3}\=\twobytwo{1}{0}{0}{-1}\~,
\label{sigmas01}
\eeq
form the algebra \e{paraquaternions} of {\em para-quaternions}, \aka the
algebra of {\em split quaternions}.

\begin{observation}
The $gl(2)$-generators $t\q_{\alpha}$, $\alpha\in\{0,1,2,3\}$,
yields a representation of $\Id$, $P$, $J$ and $\Sigma$, respectively.
\end{observation}

\noi
Moreover, recall that the adjoint representation of $SL(2)$, which acts on the
$sl(2)$-generators $t\q_{\alpha}$ by conjugation, is isomorphic to the
restricted Lorentz group $SO^{+}(2,1)$ in $2\!+\!1$ dimensions. This implies
that the para-hypercomplex structure $\{ P; J; \Sigma\}$ implements a global
$O(2,1)$ Lorentz symmetry. See Appendix~\ref{appliegroup} for further details.

\noi
A $GL(2)$ rotation \e{gl2rotationpb} of the Poisson brackets
$\{\cdot,\cdot\}^{a}$ induces an action
``.'' :$GL(2)\times C^{\infty}(\cN) \to C^{\infty}(\cN)$ on the Casimir variables
\beq
\xi\q_{ia}\~\to\~\xi^{\prime}_{jb} \= \xi^{\prime}_{jb}(g,\xi)\~,
\qquad g\in GL(2)\~, \qquad
\left. \xi^{\prime}_{jb}\right|_{g={\bf 1}\q_{2 \times 2}} \=\xi\q_{jb}\~,
\label{gl2rotationxi}
\eeq
such that $\{\cdot,\xi^{\prime}_{ai}\}^{\prime b}$ stays diagonal in the
$\q_{a}{}^{b}$ indices. We stress that the action \e{gl2rotationxi} is in
general {\em not} given by a linear $GL(2)$-rotation 
$\xi\q_{ai}\to\xi^{\prime}_{ai}=g\q_{a}{}^{b}\~\xi\q_{bi}$, although it is indeed
the case in bi-Darboux coordinates, \cf \eq{gl2rotationpee}.
The Casimir variables $\xi\q_{ia}$ are in general a sort of generalized $GL(2)$
doublets in the sense of \eq{gl2rotationxi}, while the fiber variables $q^i$
are genuine $GL(2)$ singlets (=invariants).

\subsubsection{Factorizable Case}
\label{secfacgroupaction}

\noi
In the factorizable case, there exists an atlas of local bi-Darboux coordinate
systems \e{bidarbouxform01}, \cf Theorem~\ref{theoremfactorization}.
In local bi-Darboux coordinates $\{q^{i}; p\q_{aj}\}$,
the globally defined structures $\{\Id; P; J; \Sigma \}$ become
\beq
\Id\=\partial^{aj}_{r}\~(t\q_{0})\q_{a}{}^{b}
\otimes\larrow{\dd p\q_{bj}}\~, \quad
P\=\partial^{aj}_{r}\~(t\q_{1})\q_{a}{}^{b}
\otimes\larrow{\dd p\q_{bj}}\~, \quad
J\=\partial^{aj}_{r}\~(t\q_{2})\q_{a}{}^{b}
\otimes\larrow{\dd p\q_{bj}}\~, \quad
\Sigma\=\partial^{aj}_{r}\~(t\q_{3})\q_{a}{}^{b}
\otimes\larrow{\dd p\q_{bj}}\~.\label{sigmapjidbidarboux}
\eeq
The formulas \e{sigmapjidbidarboux} are invariant under bi-canonical
transformations, \cf Section~\ref{secbict}. 

\noi
A $GL(2)$ rotation \e{gl2rotationpb} of the Poisson brackets
$\{\cdot,\cdot\}^{a}$ corresponds to a $GL(2)$ rotation of the momenta
\beq
p\q_{ai}\~\to\~p^{\prime}_{ai} \= g\q_{a}{}^{b}\~p\q_{bi}\~ \~, 
\qquad g\in GL(2)\~, 
\label{gl2rotationpee}
\eeq
here written as a left group action. 
The $GL(2)$ rotation \e{gl2rotationpee} induces a conjugation 
$t\q_{\alpha}\to g^{-1}t\q_{\alpha}g$ of the $gl(2)$-generators $t\q_{\alpha}$
in \eq{sigmapjidbidarboux}, which in turn leads to a restricted Lorentz
transformation of the para-hypercomplex structure $\{ P; J; \Sigma\}$,
\cf Appendix~\ref{appliegroup}.

\vspace{0.8cm}

\noi
{\sc Acknowledgement:}~
K.B.\ would like to thank M.~Vasiliev and the Lebedev Physics Institute for 
warm hospitality. The work of I.A.B.\ is supported by grants RFBR 11--01--00830
and RFBR 11--02--00685. The work of K.B.\ is supported by the Grant agency 
of the Czech republic under the grant P201/12/G028.
\appendix

\section{Bi-Poincar\'e Lemma}
\label{appbipoincarelemma}

\subsection{Algebra $\cA$}

\noi
Consider $3n$ coordinates $x^{i}_{\alpha}$, $i\in\{1,\ldots,n\}$, 
$\alpha\in\{1,2,3\}$, that are defined in an open neighborhood of the origin, 
and with Grassmann parity $\eps(x^{i}_{\alpha})=\eps_{i}+\delta^{3}_{\alpha}$. 
Define three integer gradings
\beq
\deg\q_{\alpha}(x^{i}_{\beta})\~:=\~\delta_{\alpha\beta}\~, \qquad
\deg\q_{\alpha}(fg)\=\deg\q_{\alpha}(f)+\deg\q_{\alpha}(g)\~,\qquad 
f,g\in\C[[x]]\~, \qquad \alpha, \beta \in\{1,2,3\}\~,
\eeq 
and three number operators
\beq
N\q_{\alpha}\~:=\~x^{i}_{\alpha}\papal{x^{i}_{\alpha}}\~,
\qquad \alpha \in\{1,2,3\}\~.
\label{ndef123}
\eeq
(No sum over $\alpha$ in the last \eq{ndef123}.) We will often refer to the 
$x^{i}_{3}$ variables as ``one-forms'', and the third grading 
``$\deg\q_{3}$'' as ``form-degree''. Let 
\beq
\cA\~:=\~\C[[x]] \= \bigoplus_{n\q_{1},n\q_{2},n\q_{3}\in\N\q_{0}}
\cA\q_{n\q_{1},n\q_{2},n\q_{3}}\~, \qquad
\cA\q_{n\q_{1},n\q_{2},n\q_{3}}\~:=\~\{\omega\in\cA \mid 
\forall \alpha \in\{1,2,3\} \~:\~\deg\q_{\alpha}(\omega)\=n\q_{\alpha}\}\~, 
\label{algebraca}
\eeq
be the algebra of formal power series in the $x$ variables. A power series 
$\omega=\omega(x)$ of the algebra $\cA$ will often be referred to as a 
``form''.

\subsection{Differentials $d^{a}$, $i\q_{a}$ and $\cL^{a}_{b}$}

\noi
Define $2$ nilpotent and commuting Grassmann-odd differentials 
\beq
d^{a}\~:=\~x^{i}_{3} \papal{x^{i}_{a}}\~, \qquad \eps(d^{a})\=1\~, 
\qquad [d^{a},d^{b}]\=0\~, \qquad a,b\in\{1,2\}\~.
\eeq
Define $2$ dual nilpotent and commuting Grassmann-odd differentials 
\beq
i\q_{a}\~:=\~x^{i}_{a} \papal{x^{i}_{3} }\~, \qquad \eps(i\q_{a})\=1\~, 
\qquad [i\q_{a},i\q_{b}]\=0\~, \qquad 
a,b\in\{1,2\}\~.
\eeq
Define their $2\times 2=4$ mutual super-commutators
\beq
\cL^{a}_{b}\~:=\~[i\q_{b},d^{a}]\= x^{i}_{b} \papal{x^{i}_{a}}
+\delta^{a}_{b}\~N\q_{3}\~, \qquad \eps(\cL^{a}_{b})\=0\~, 
\qquad a,b\in\{1,2\}\~. 
\eeq
Explicitly, they are
\beq
\cL^{1}_{1}\=N\q_{1}+N\q_{3}\~, \qquad 
\cL^{2}_{2}\=N\q_{2}+N\q_{3}\~, \qquad 
\cL^{2}_{1}\= x^{i}_{1} \papal{x^{i}_{2}}\~, \qquad
\cL^{1}_{2}\= x^{i}_{2} \papal{x^{i}_{1}}\~.
\eeq
In particular, define the trace
\beq
\cL\~:=\~\cL^{a}_{a}\=\cL^{1}_{1}+\cL^{2}_{2}
\=N\q_{1}+N\q_{2}+2N\q_{3}\~.\label{apptrace}
\eeq
The $\cL^{a}_{b}$ operators form a $gl(2,\C)$ Lie algebra,
\beq
[\cL^{a}_{b},\cL^{c}_{d}]\=\delta^{a}_{d}\~\cL^{c}_{b}
-\delta^{c}_{b}\~\cL^{a}_{d}\~,\qquad a,b,c,d\in\{1,2\}\~.
\eeq
The following formulas hold
\beq
[\cL^{a}_{c},d^{b}]\=\delta^{a}_{c}\~ d^{b}-(a \leftrightarrow b)\~, \qquad
[i\q_{a},\cL^{c}_{b}]\=i\q_{a}\~\delta^{c}_{b}-(a \leftrightarrow b)\~,
\qquad a,b,c\in\{1,2\}\~.
\eeq

\subsection{$d$ and $i$}

\noi
Define also nilpotent second-order differential operators  
\beq
d\~:=\~\Hf \epsilon\q_{ba}\~d^{a}d^{b}\=d^{1}d^{2}\~, 
\qquad i\~:=\~\Hf \epsilon^{ab}\~i\q_{b}i\q_{a}\=i\q_{2}i\q_{1}\~,
\qquad \eps(d)\=0\~,  \qquad \eps(i)\=0\~.
\eeq
Here the sign convention for the Levi-Civita $\epsilon$-tensor is 
\beq 
\epsilon^{ab}\~\epsilon\q_{bc}\=\delta^{a}_{c}\~, \qquad 
\epsilon^{12}\=\epsilon_{21}\=+1\~.
\eeq
The following formulas hold
\beq
[\cL^{a}_{b},d]\=\delta^{a}_{b}\~d\~, \qquad
[i,\cL^{a}_{b}]\=\delta^{a}_{b}\~i\~, \qquad  a,b\in\{1,2\}\~,
\eeq
\beq
[\cL,d]\=2d\~,\qquad [i,\cL]\=2i\~, \label{tracedi}
\eeq
\beq
[d^{a},i]
\=(\cL^{a}_{c}+\delta^{a}_{c})\~ i\q_{b}\~\epsilon^{bc}
\=\cL^{a}_{c}\~ i\q_{b}\~\epsilon^{bc}+i\q_{b}\~\epsilon^{ba}\~,
\qquad a\in\{1,2\}\~. 
\eeq

\subsection{$sl(2,\C)$ Lie Algebra}

\noi
We now decompose the four-dimensional Lie algebra 
$gl(2,\C)=\C\oplus sl(2,\C)$. The trace operator 
$\cL$ is the generator of the center $\C$.
The three $sl(2,\C)$ generators $J\q_{\alpha}$, $\alpha\in\{1,2,3\}$, 
are defined as some linear combinations of the four $gl(2,\C)$ 
generators $\cL^{a}_{b}$, $a,b\in\{1,2\}$,
\beq
J\q_{1}\~:=\~\frac{\cL^{2}_{1}+\cL^{1}_{2}}{2}\~, \qquad 
J\q_{2}\~:=\~\frac{\cL^{2}_{1}-\cL^{1}_{2}}{2\mathrm{i}}\~, \qquad 
J\q_{3}\~:=\~\frac{\cL^{1}_{1}-\cL^{2}_{2}}{2}
\= \frac{N\q_{1}-N\q_{2}}{2}\~, \label{jay01}
\eeq
\beq
J\q_{\pm}\=J\q_{1}\pm \mathrm{i} J\q_{2}\~,\qquad
J\q_{+}\=\cL^{2}_{1}\~,\qquad J\q_{-}\=\cL^{1}_{2}\~,\qquad
J^{2}\= J^{2}_{1}+ J^{2}_{2}+ J^{2}_{3}\~.\label{jay02}
\eeq
It is straightforward to check that $J\q_{\alpha}$, $\alpha\in\{1,2,3\}$, form
a $sl(2,\C)$ Lie algebra,
\beq
[J\q_{\alpha},J\q_{\beta}]
\=\mathrm{i}\epsilon\q_{\alpha\beta\gamma}\~J\q_{\gamma}\~, \qquad 
\epsilon\q_{123}\=+1\~, \qquad \alpha, \beta, \gamma \in\{1,2,3\}\~.
\eeq
Several operators commute with the $sl(2,\C)$ generators
$J\q_{\alpha}$, $\alpha\in\{1,2,3\}$,
\beq
[J^{2},J\q_{\alpha}]\=0\~, \quad
[\cL,J\q_{\alpha}]\=0\~, \quad
[N\q_{3},J\q_{\alpha}]\=0\~, \quad 
[d,J\q_{\alpha}]\=0\~,\quad
[i,J\q_{\alpha}]\=0\~, \quad 
\alpha\in\{1,2,3\}\~.
\label{sl2casimirs}
\eeq

\subsection{Bi-Poincar\'e Lemma}

\begin{lemma}[Bi-Poincar\'e Lemma]
A $d^{a}$-closed form $\omega=\omega(x)$, $a\in\{1,2\}$, that does not contain
zero- and one-forms, is locally $d$-exact. Or equivalently, in symbols: 
\beq
\forall \omega\in\cA\~:\qquad \left\{ \begin{array}{rcl} 
\forall a\in\{1,2\}&:&(d^{a}\omega)\=0 \cr\cr
\deg\q_{3}(\omega) &\geq& 2 \end{array} \right\} 
\qquad \Rightarrow  \qquad\exists \eta\in\cA\~:\~ \omega\=(d\eta)\~.
\label{formulabipoincarelemma}
\eeq
\label{lemmabipoincarelemma}
\end{lemma}

\noi
Here we have defined the $\cA$ algebra \e{algebraca} to be the algebra 
$\C[[x]]$ of formal power series with complex coefficients. By decomposing
\eq{formulabipoincarelemma} in real and imaginary parts, it is clear that the
bi-Poincar\'e Lemma~\ref{lemmabipoincarelemma} also holds if one instead
considers the algebra $\R[[x]]$ of formal power series with real coefficients. 

\noi
Because the set $\{x^{1}_{1},\ldots,x^{n}_{1} ;x^{1}_{2},\ldots, x^{n}_{2}\}$ is
twice as big as the set $\{x^{1}_{3},\ldots,x^{n}_{3}\}$, we cannot apply the
proof technique of \eg \Ref{bltha2} and \Ref{bltla2}, which requires a 
balanced number of variables. Instead we will use a bit of $sl(2,\C)$
representation theory to obtain the pertinent estimate \e{crucial}.

\subsection{$L$ and $\Lambda$}

\noi
Define a third-order differential operator
\beq
L\~:=\~[d,i]\=\Hf\epsilon\q_{ba}[d^{a}d^{b},i]
\=\Hf \epsilon\q_{ba}[d^{a},[d^{b},i]] + \epsilon\q_{ba}[d^{a},i]d^{b}
\= \Lambda + R\q_{b}\~d^{b}\~,
\label{appdefl}
\eeq
where
\beq
R\q_{b}\~:=\~\epsilon\q_{ba}\~\cL^{a}_{c}\~ i\q_{d}\~\epsilon^{dc}
\~,\qquad b\in\{1,2\}\~,
\eeq
and where
\beq
\Lambda\~:=\~ -\Hf \epsilon\q_{ba}\~\cL^{a}_{d}\~\cL^{b}_{c}\~\epsilon^{cd}
-\frac{\cL}{2}
\= \Hf \{\cL^{1}_{1},\cL^{2}_{2}\}\q_{+}
-\Hf \{\cL^{2}_{1},\cL^{1}_{2}\}\q_{+} -\frac{\cL}{2}
\=\frac{\cL}{2}\left(\frac{\cL}{2}-1\right)-J^{2}\~.
\label{appdeflambda}
\eeq
To prove the last equality in \eq{appdeflambda}, note that
\beq
\Hf \{\cL^{1}_{1},\cL^{2}_{2}\}\q_{+}
\=\left(\frac{\cL}{2}\right)^{2} - J^{2}_{3}\~, \qquad  
\Hf \{\cL^{2}_{1},\cL^{1}_{2}\}\q_{+}\=\Hf \{J\q_{+}, J\q_{-}\}\q_{+}
\= J^{2}_{1}+ J^{2}_{2} \~. 
\eeq
The operators $L, \Lambda\in\End(\cA)$ are $gl(2,\C)$ Casimirs,
\beq
[\cL^{a}_{b},L]\=0\~,\qquad [\cL^{a}_{b},\Lambda]\=0\~, \qquad a,b\in\{1,2\}\~.
\label{gl2casimirs}
\eeq
Since $\Lambda$ is a quadratic polynomial \e{appdeflambda} of the four
$gl(2,\C)$ generators $\cL^{a}_{b}$, it follows from \eq{gl2casimirs}
that $L$ and $\Lambda$ commute
\beq
[L,\Lambda]\=0\~. \label{llambdacom}
\eeq

\subsection{Zero-Modes for $\Lambda$?}

\noi
Define for later convenience
\beq
\Lambda^{\prime}\~:=\~\left. \Lambda\right|\q_{\cL\to\cL - 2}\=
\left(\frac{\cL}{2}-1\right)\left(\frac{\cL}{2}-2\right)-J^{2}\~,
\eeq 
so that 
\beq
d\~f(\Lambda)\=f(\Lambda^{\prime})\~d\~, \qquad 
f(\Lambda)\~i\=i\~f(\Lambda^{\prime})\~,
\label{appconvenient}
\eeq
where $f=f(\lambda)$ is some function of $\lambda\in\C$, \cf 
\eqs{tracedi}{sl2casimirs}.

\begin{lemma}
\bea
\ker\Lambda\~\cap\~
\{\omega\in\cA \mid \deg\q_{3}(\omega)\~\geq\~2\} &=& \{0\}\~, 
\label{lambdainv01} \\ 
\ker\Lambda^{\prime}\~\cap\~
\{\omega\in\cA \mid \deg\q_{3}(\omega)\~\geq\~4\} &=& \{0\}\~.
\label{lambdainv02} 
\eea 
\label{lemmalambdainv}
\end{lemma}

\noi
{\sc Proof of Lemma~\ref{lemmalambdainv}}:\~\~We will only here prove the first
statement \e{lambdainv01}, as the second statement \e{lambdainv02} is similar.
The vector space $\cA$ becomes an infinite-dimensional representation of
$sl(2,\C)$ by acting with the $J\q_{\alpha}$ generators \e{jay01} from
the left. Since $N\q_{12}:=N\q_{1}+N\q_{2}=\cL-2N\q_{3}$ and $N\q_{3}$ are
$sl(2,\C)$ Casimirs, we only have to consider a finite-dimensional
subspace 
\beq
\cA\q_{n\q_{12},n\q_{3}}\= \{\omega\in\cA \mid 
\deg\q_{1}(\omega)+\deg\q_{2}(\omega)\=n\q_{12} 
\~\wedge\~ \deg\q_{3}(\omega)\=n\q_{3}\}\~,
\eeq
for a pair of non-negative integers 
$n\q_{12}\in\N\q_{0}\equiv\{0,1,2,\ldots\}$ and
$n\q_{3}\in\{2,3,4,\ldots\}$. 
The eigenvalue $\ell$ of the trace operator $\cL=N\q_{12}+2N\q_{3}$ inside 
$\cA\q_{n\q_{12},n\q_{3}}$ is
\beq
\ell\=~n\q_{12}+2n\q_{3}\~.
\eeq
The two number operators $N\q_{1}$ and $N\q_{2}$ are diagonalizable 
inside the pertinent subspace
\beq 
\cA\q_{n\q_{12},n\q_{3}}
\=\bigoplus_{n\q_{1},n\q_{2}\in\N\q_{0}}^{n\q_{1}+n\q_{2}=n\q_{12}}
\cA\q_{n\q_{1},n\q_{2},n\q_{3}}
\eeq
with bounded eigenvalues
\beq
n\q_{1},n\q_{2}\~\in\~\{0,1,2,\ldots,n\q_{12}\}\~. \label{n1n2bounded}
\eeq 
According to (a superversion of) Weyl's Theorem\footnote[1]{It is possible to
explicitly construct a sesqui-linear form
$\langle\cdot,\cdot\rangle:\cA\times\cA\to\C$ that turns (the representation of)
the generators $J\q_{\alpha}$, $\alpha\in\{1,2,3\}$, into Hermitian operators.
This is the setting of Weyl's Theorem often stated in the Physics literature.
However, Weyl's Theorem does actually not rely on the existence of any
Hermitian structure, see \eg \Ref{humphreys72}.},
a finite-dimensional representation $\cA\q_{n\q_{12},n\q_{3}}$ of a semisimple Lie
algebra is always {\em completely reducible}, \ie a finite direct sum of irreps
(=irreducible representations) 
\beq
\cA\q_{n\q_{12},n\q_{3}} 
\= \bigoplus_{j\in \Hf \N\q_{0}} \mu\q_{j} V\p_{j}\~. \label{appasum}
\eeq
Here $\mu\q_{j}\in\N\q_{0}$ denotes the multiplicity, \ie how many
times the $(2j\!+\!1)$-dimensional irrep $V\p_{j}$ appears in the direct sum
\e{appasum}, where $j\in \Hf \N\q_{0}$ is a non-negative half-integer.
Recall that the eigenvalues of $J^{2}$ and $J\q_{3}$ on $V\p_{j}$ are
\beq
j(j+1) \qquad {\rm and} \qquad  m\~\in\~\{-j,1\!-\!j,\ldots,j\!-\!1,j\}\~, 
\eeq
respectively. Since $\Lambda$ is a Casimir, the irrep $V\p_{j}$ is an
eigenspace for $\Lambda$ with some eigenvalue $\lambda$, \cf Schur's Lemma. In
particular, the operator $\Lambda$ is diagonalizable on the full vector space
$\cA$. We have to show that there are no zero eigenvalues $\lambda\neq 0$.
Inside $V\p_{j}\subseteq\cA\q_{n\q_{12},n\q_{3}}$, the eigenvalues $m$ for
$J\q_{3}=\Hf(N\q_{1}-N\q_{2})$ must satisfy $|m|\leq\Hf n\q_{12}$, \cf 
\eq{n1n2bounded}.
In particular, this must be true for the largest eigenvalue $m=j$. Hence 
\beq
j\~\leq\~\frac{n\q_{12}}{2}\~. \label{crucial}
\eeq
Therefore
\bea
\lambda&\equi{\e{appdeflambda}}&
\frac{\ell}{2}\left(\frac{\ell}{2}-1\right)-j(j+1) \cr
&\geq&
\left(\frac{n\q_{12}}{2}+n\q_{3}\right)\left(\frac{n\q_{12}}{2}+n\q_{3}-1\right)
-\frac{n\q_{12}}{2}\left(\frac{n\q_{12}}{2}+1\right)   
\= (n\q_{12}+n\q_{3})(n\q_{3}-1)\~>0\~,
\eea
because $n\q_{12}\geq 0$ and $n\q_{3}\geq 2$. In particular, the operator 
$\Lambda$ is strictly positive. 
\proofbox

\subsection{Proof of Bi-Poincar\'e Lemma~\ref{lemmabipoincarelemma}}

\noi
{\sc Proof of Bi-Poincar\'e Lemma~\ref{lemmabipoincarelemma}}:\~\~Let there 
be given a $d^{a}$-closed form $\omega\in \cA$ with $\deg\q_{3}(\omega)\geq 2$.
Define a form
\beq
\eta\~:=\~(i\Lambda^{-1}\omega)
\~\equi{\e{appconvenient}}\~(\Lambda^{\prime -1}i\omega)\~,\label{appdefeta}
\eeq
which is well-defined because of Lemma~\ref{lemmalambdainv}.
Then we calculate
\bea
(d\eta)&\equi{\e{appdefeta}}&(di\Lambda^{-1}\omega)
\~\equi{\e{appdefl}}\~(L\Lambda^{-1}\omega)+(id\Lambda^{-1}\omega)
\~\equi{\e{llambdacom}}\~(\Lambda^{-1}L\omega)+(id\Lambda^{-1}\omega) \cr
&\equi{\e{appdefl}+\e{appconvenient}}&
(\Lambda^{-1}(\Lambda + R\q_{b}\~d^{b})\omega) 
+ (i\Lambda^{\prime -1} d\omega)
\~\equi{\omega\~{\rm closed}}\~\omega \~. 
\eea
\proofbox

\section{Real Lie Groups}
\label{appliegroup}

\noi
Here we collect some facts about the real Lie Groups $SO^{+}(2,1)$, $SL(2)$ 
and $GL(2)$ used in the main text.

\subsection{$SO^{+}(2,1)$}
\label{seclorentz}

\noi
Let the {\em Minkowski metric} in $2\!+\!1$ real dimensions be 
\beq
\eta\q_{\alpha\beta}\~:=\~\threebythree{1}{0}{0}{0}{-1}{0}{0}{0}{1}\~.
\label{minkowski}
\eeq
(The non-standard ordering of spatial and temporal directions in the metric 
\e{minkowski} is related to that the $\sigma\q_{2}$ Pauli matrix \e{sigmas02} 
is imaginary, \cf \eq{paraquaternions}.)
The {\em Lorentz group} is 
\beq
O(2,1):=\{\Lambda \in \Mat\q_{3\times 3}(\R)\mid 
\Lambda^{T}\eta \Lambda =\eta\}\~.
\eeq
The {\em restricted Lorentz group} is
\beq
SO^{+}(2,1)\~:=\~\{\Lambda \in \Mat\q_{3\times 3}(\R)\mid 
\Lambda^{T}\eta \Lambda =\eta \~\wedge\~ \det(\Lambda)=1 \~\wedge\~ 
\Lambda^{2}\q_{2}\geq 1 \}\=e^{so(2,1)}\~,
\label{lorentzgroup21}
\eeq
and its Lie algebra
\beq
so(2,1)\~:=\~ \{\lambda\in \Mat\q_{3\times 3}(\R)\mid 
\lambda^{T}=-\eta \lambda\eta^{-1} \}
\={\rm span}\q_{\R} \{T\q_{\alpha} \mid \alpha \in\{1,2,3\}\}\~,
\eeq
with generators $T\q_{\alpha}$, $\alpha\in\{1,2,3\}$, satisfying
\beq
[T\q_{\alpha}, T\q_{\beta}]
\=\sqrt{-\eta}\~\epsilon\q_{\alpha\beta\gamma}\~\eta^{\gamma\delta}\~T\q_{\delta}\~,
\qquad \alpha, \beta, \gamma, \delta \in\{1,2,3\}\~. 
\eeq
Here
\beq
\eta\~:=\~\det(\eta\q_{\alpha\beta})\=-1
\eeq
is the determinant of the Minkowski metric $\eta\q_{\alpha\beta}$. 
One may, \eg choose generators
\beq
(T\q_{\alpha})^{\delta}\q_{\beta} 
\=\sqrt{-\eta}\~\epsilon\q_{\alpha\beta\gamma}\~\eta^{\gamma\delta} \~,
\qquad \alpha, \beta, \gamma, \delta \in\{1,2,3\}\~, 
\eeq
so that
\beq
T\q_{1}\= \threebythree{0}{0}{0}{0}{0}{1}{0}{1}{0}\~,\qquad
T\q_{2}\= \threebythree{0}{0}{1}{0}{0}{0}{-1}{0}{0}\~,\qquad
T\q_{3}\= \threebythree{0}{-1}{0}{-1}{0}{0}{0}{0}{0}\~.
\eeq
$T\q_{1}$ and $T\q_{3}$ generate Lorentz boosts, while $T\q_{2}$ generates
spatial rotations. The Levi-Civita $\epsilon\q_{\alpha\beta\gamma}$-symbol
satisfies
\beq
\eta\~\epsilon\q_{\alpha\beta\mu}\~\eta^{\mu\nu}\~\epsilon\q_{\nu\gamma\delta} 
\= \eta\q_{\alpha\gamma}\~\eta\q_{\beta\delta}
-\eta\q_{\alpha\delta}\~\eta\q_{\beta\gamma}\~,\qquad  
\epsilon\q_{123}\=+1\~.
\eeq

\subsection{$SL(2)$}
\label{secsl2}

\noi
The {\em special linear group} in $2$ real dimensions is
\beq
SL(2)\~:=\~\{g\in \Mat\q_{2\times 2}(\R) \mid \det(g)=1 \}
\=Sp(2)\=e^{sl(2)}\~, 
\eeq
and its Lie algebra
\beq
sl(2)\~:=\~ \{x\in \Mat\q_{2\times 2}(\R) \mid \tr(x)=0 \} 
\={\rm span}\q_{\R} \{t\q_{\alpha} \mid \alpha \in\{1,2,3\}\}\~, 
\eeq
with generators $t\q_{\alpha}$, $\alpha\in\{1,2,3\}$, satisfying
\beq
t\q_{\alpha}t\q_{\beta}\= \eta\q_{\alpha\beta}{\bf 1}\q_{2\times 2}
+\sqrt{-\eta}\~\epsilon\q_{\alpha\beta\gamma}\~\eta^{\gamma\delta}\~t\q_{\delta}\~,
 \qquad \alpha, \beta, \gamma, \delta \in\{1,2,3\}\~,  \label{paraquaternions}
\eeq
where $\eta\q_{\alpha\beta}$ is the Minkowski metric \e{minkowski} in 
$2\!+\!1$ dimensions. One may, \eg choose generators
\beq
t\q_{1}\~:=\~\sigma\q_{1}\~:=\~\twobytwo{0}{1}{1}{0}\~, \qquad 
t\q_{2}\~:=\~ -\mathrm{i}\sigma\q_{2}\~:=\~\twobytwo{0}{-1}{1}{0}\~, \qquad 
t\q_{3}\~:=\~\sigma\q_{3}\~:=\~\twobytwo{1}{0}{0}{-1}\~.
\label{sigmas02}
\eeq
Here $\sigma\q_{\alpha}$, $\alpha\in\{1,2,3\}$, are the Pauli matrices,
which satisfies 
\beq
\sigma\q_{\alpha}\sigma\q_{\beta}\= \delta\q_{\alpha\beta}{\bf 1}\q_{2\times 2}
+\mathrm{i}\epsilon\q_{\alpha\beta\gamma}\~\sigma\q_{\gamma}\~,\qquad 
\alpha, \beta, \gamma \in\{1,2,3\}\~.  
\eeq

\subsection{$GL(2)$}
\label{secgl2}

\noi
The {\em general linear group} in $2$ real dimensions is
\beq
GL(2)\~:=\~\{g\in \Mat\q_{2\times 2}(\R) \mid \det(g)\neq 0\}\=e^{gl(2)}
\~\cong\~\R^{\times} \times SL(2)\~,  \qquad  
\R^{\times}\~:=\~\R \backslash \{0\}\~,
\eeq
where the Abelian factor $\R^{\times}$ stores the value of the determinant
$\det(g)$.  The corresponding Lie algebra of $GL(2)$ is
\beq
gl(2)\=\Mat\q_{2\times 2}(\R)\=\End(\R^{2})
\=Z(gl(2)) \oplus sl(2)\~\cong\~\R \oplus sl(2)\~.
\eeq
The Lie group and Lie algebra centers are
\beq
Z(GL(2))\=\R^{\times}t\q_{0}\~, \qquad
Z(gl(2))\=\R t\q_{0}\~, \qquad 
t\q_{0}\~:=\~{\bf 1}\q_{2\times 2}\~.
\eeq
The $gl(2)$-generators $t\q_{\alpha}$, $\alpha\in\{0,1,2,3\}$, form 
the algebra \e{paraquaternions} of {\em para-quaternions}, \aka the algebra
of {\em split quaternions}.

\subsection{$SO^{+}(2,1)\~\cong\~\Ad(SL(2))$}
\label{secso21isadsl2}

\begin{observation}
The real Lie algebras $so(2,1)$ and $sl(2)$ are isomorphic
$so(2,1)\~\cong\~sl(2)$ via the map $T\q_{\alpha}\mapsto \Hf t\q_{\alpha}$.
\end{observation}

\noi
Recall that the {\em adjoint Lie group representation} 
$\Ad: SL(2)\to\End(sl(2))$ and {\em the adjoint Lie algebra representation} 
$\ad: sl(2)\to\End(sl(2))$ are defined as
\beq
\Ad(g)x \~:= gxg^{-1}\~,\qquad  g\in SL(2)\~, \qquad x\in sl(2)\~, 
\eeq
and 
\beq 
  \ad(x)y \= [x,y]\~,\qquad x,y\~\in\~  sl(2)\~,
\eeq
respectively.

\noi
The Lie algebra $sl(2)$ may be identified with Minkowski space
$M(2,1)\cong sl(2)$ because the determinant is the the Minkowski metric (up to
a sign),
\beq
\det(x)
\=\det \twobytwo{x^{3}}{x^{1}-x^{2}}{x^{1}-x^{2}}{-x^{3}}
\=-x^{\alpha}\~\eta\q_{\alpha\beta}\~x^{\beta}\~,  
\qquad x\=x^{\alpha}t\q_{\alpha}\~\in\~sl(2)\~:=\~\{x \mid \tr(x)=0\}\~.
\eeq
Since the conjugation $Ad(g)x$ with an element $g\in GL(2)$ preserves traces
and determinants, and hence Minkowski lengths, the group element $g$ must
correspond to a Lorentz transformation $\Lambda\in O(2,1)$ of the Minkowski
space $M(2,1)$. The following Proposition~\ref{so21isadsl2} is a refinement of
this fact.

\begin{proposition} 
The restricted Lorentz group $SO^{+}(2,1)$ is isomorphic to the adjoint
representation of $SL(2)$,
\beq
 SO^{+}(2,1)\~\cong\~\Ad(SL(2))\~\cong\~SL(2) / \Z\q_{2}\~. 
\label{so21isadsl2a}
\eeq
The Lie group isomorphism is given by the map
\beq
\Ad(e^{\Hf x^{\alpha} t\q_{\alpha}})t\q_{\beta}
\= e^{\Hf x^{\alpha} \ad(t\q_{\alpha}) }t\q_{\beta} 
\= t\q_{\alpha}\left(e^{x^{\gamma} T\q_{\gamma} }\right)^{\alpha}\q_{\beta}\~, \qquad
x^{\alpha}\in\R\~. 
\label{so21isadsl2b} 
\eeq
\label{so21isadsl2}
\end{proposition}

\noi
In particular, $SL(2)$ is a double cover of $SO^{+}(2,1)$, because
$\Ad(\pm {\bf 1}\q_{2\times 2})={\bf 1}\q_{3\times 3}$.
Equation \e{so21isadsl2b} says in words that conjugating an $sl(2)$-generator
$t\q_{\alpha}$ with a $SL(2)$ matrix $g=e^{\Hf x^{\alpha} t\q_{\alpha}}$ corresponds
to a restricted Lorentz transformation $\Lambda=e^{ x^{\alpha} T\q_{\alpha}}$ of the
three $sl(2)$-generators $t\q_{\alpha}$. 
The last equality in \eq{so21isadsl2b} can, \eg be proved by scaling the
variable $x^{\alpha}\to rx^{\alpha}$ with a radial $1$-parameter $r\geq 0$, and
show that the \lhs and the \rhs satisfy the same first-order ODE \wrtt radial 
parameter $r$, and same initial condition at $r=0$.


\begin{thebibliography}{999}

\bibitem{magri78}
{}F.~Magri, J.~Math.~Phys.\ {\bf 19} (1978) 1156. 

\bibitem{gelfand00} I.M.~Gelfand and I.~Zakharevich, 
Selecta Math.\ {\bf 6} (2000) 131, arXiv:math.DG/9903080.

\bibitem{bltla2} I.A.~Batalin, P.~Lavrov and I.~Tyutin, 
J.~Math.~Phys.\ {\bf 31} (1990) 1487; 
{\em ibid.} {\bf 32} (1991) 532; 
{\em ibid.} {\bf 32} (1991) 2513.

\bibitem{bm95} I.A.~Batalin and R.~Marnelius, 
Phys.~Lett.\ {\bf B350} (1995) 44, arXiv:hep-th/9502030.

\bibitem{bms95} I.A.~Batalin, R.~Marnelius and A.M.~Semikhatov,
Nucl.~Phys.\ {\bf B446} (1995) 249, arXiv:hep-th/9502031.

\bibitem{djb95} P.H.~Damgaard, F.~De Jonghe and K.~Bering,
Nucl.~Phys.\ {\bf B455} (1995) 440, arXiv:hep-th/9508034.

\bibitem{bm96} I.A.~Batalin and R.~Marnelius,
Nucl.~Phys.\ {\bf B465} (1996) 521, arXiv:hep-th/9510201.

\bibitem{bv81} 
I.A.~Batalin and G.A.~Vilkovisky, Phys.~Lett.\ {\bf 102B} (1981) 27.

\bibitem{bv83} I.A.~Batalin and G.A.~Vilkovisky,
Phys.~Rev.\ {\bf D28} (1983) 2567 [E: {\bf D30} (1984) 508].

\bibitem{bv84} 
I.A.~Batalin and G.A.~Vilkovisky, Nucl.~Phys.\ {\bf B234} (1984) 106.

\bibitem{bv85} 
I.A.~Batalin and G.A.~Vilkovisky, J.~Math.~Phys.\ {\bf 26} (1985) 172. 

\bibitem{grisemi97} M.A.~Grigoriev and A.M.~Semikhatov, 
Phys.~Lett.\ {\bf B417} (1998) 259, arXiv:hep-th/9708077.

\bibitem{grisemi98} M.A.~Grigoriev and A.M.~Semikhatov, 
Theor.~Math.~Phys.\ {\bf 124} (2000) 1157, arXiv:hep-th/9807023.

\bibitem{obata56} M.~Obata, Japan J.~Math.\ {\bf 26} (1956) 43.

\bibitem{swann05}
A.S.~Dancer, H.R.~J{\o}rgensen and A.F.~Swann,
Rend.~Semin.~Mat.~Torino {\bf 63} (2005) 119, arXiv:math.DG/0412215.  

\bibitem{andrada05}
A.~Andrada, Annals~of~Global~Analysis~and~Geometry {\bf 27} (2005) 377.

\bibitem{alekseevsky08}
D.V.~Alekseevsky and V.~Cort\'{e}s, Osaka~J.~Math.\ {\bf 45} (2008) 215.

\bibitem{alekseevsky09} D.V.~Alekseevsky, C.~Medori and A.~Tomassini, 
Russian~Math.~Surveys {\bf 64} (2009) 1, arXiv:0806.2272.

\bibitem{gotelind11} M.~G\"oteman and U.~Lindstr\"om, 
Lett.~Math.~Phys.\ {\bf 95} (2011) 211, arXiv:0903.2376.

\bibitem{gri99} 
M.A.~Grigoriev, Phys.~Lett.\ {\bf B458} (1999) 499, arXiv:hep-th/9901046.

\bibitem{lm87}
P.~Libermann and C.~Marle, {\em Symplectic Geometry and Analytical Mechanics},
Springer, 1987.

\bibitem{goldstein80}
H.~Goldstein, {\em Classical Mechanics}, 2nd ed., Reading, Massachusetts,
Addison--Wesley Publishing, 1980.

\bibitem{moser65}
J.~Moser, {\em On the volume elements on a manifold}, Trans.~Amer.~Math.~Soc.\
{\bf 120} (1965) 286-294.

\bibitem{b97} K.~Bering, {\em Almost parity structure, connections and
vielbeins in BV geometry}, arXiv:physics/9711010.

\bibitem{bltha2}
I.A.~Batalin, P.M.~Lavrov and I.V.~Tyutin,
J.~Math.~Phys.\  {\bf 31} (1990) 6; 
{\em ibid.}~{\bf 31} (1990) 2708;
Int.~J.~Mod.~Phys.\ A {\bf 6} (1991) 3599.

\bibitem{humphreys72}
J.E.~Humphreys, {\em Introduction to Lie Algebras and Representation Theory},
Springer, 1972.

\end{thebibliography}
\end{document}